\RequirePackage{fix-cm}
\documentclass[twocolumn]{STJour3}        	
\smartqed  

\usepackage{graphicx}
\usepackage{amsmath}
\usepackage{amsfonts}
\usepackage{amssymb}
\usepackage{subfigure}
\usepackage{graphicx}
\usepackage{epstopdf}
\usepackage{ifthen}
\usepackage{array}
\usepackage{mathabx}
\usepackage{tabularx}
\usepackage{booktabs}

\newcommand{\picdir}{.}
\newcommand{\setstyle}[1]{{\mathbb #1}}				
\newcommand{\divg}{\mathrm{div}}				
\newcommand{\EW}[1][leer]{\ensuremath{\mathbf{E}}\ifthenelse{\equal{#1}{leer}}{\;}{\ensuremath{\left[{#1} \right]}}}
\newcommand{\Var}[1][leer]{\ensuremath{\mathbf{Var}}\ifthenelse{\equal{#1}{leer}}{\;}{\ensuremath{\left[{#1} \right]}}}

\newcommand{\abs}[1]{\left|#1\right|}				
\newcommand{\prdV}[2]{\ensuremath{\left\langle #1,\,#2\right\rangle_{\Lp[2](\WR)}}} 
\newcommand{\half}{\frac{1}{2}}
\renewcommand{\ln}{\mathrm{ln\;}}					

\newcommand{\vx}{\ensuremath{\mathbf{x}}}			
\newcommand{\x}{\ensuremath{x}}					
\newcommand{\rd}{\ensuremath{r}}				
\newcommand{\rdi}[1]{\ensuremath{\rd_{#1}}}			

\newcommand{\setR}{\ensuremath{\setstyle{R}}}			
\newcommand{\setN}{\ensuremath{\setstyle{N}}}			
\newcommand{\setD}{D}						
\newcommand{\setI}{\ensuremath{\mathcal{I}}}			
\newcommand{\setIt}{\ensuremath{\tilde{\setI}}}			
\newcommand{\Lp}[1][p]{\ensuremath{L^{#1}}}			
\newcommand{\Pm}{\ensuremath{\setstyle{P}}}			
\newcommand{\Field}[1][F]{\ensuremath{\mathcal{#1}}}		
\newcommand{\WR}{\ensuremath{\Omega}}				
\newcommand{\WRM}{\ensuremath{\left(\WR,\Field,\Pm \right)}}	
\newcommand{\SE}[3]{\ensuremath{\mathcal{X}^{#2,\,#3}_{#1}}}	

\newcommand{\w}{w}						
\newcommand{\nw}{g}						
\newcommand{\coo}{\ensuremath{\mathrm{CO}_2}}			
\newcommand{\nfx}[1]{\ensuremath{H_{#1}}}			
\newcommand{\fx}{f}						
\newcommand{\fxa}[1]{\ensuremath{\fx_{#1}}}			
\newcommand{\fxg}{\fxa\nw}					
\newcommand{\mobi}[1]{\ensuremath{\lambda_{#1}}}		
\newcommand{\mn}{\mobi{\nw}}					
\newcommand{\mw}{\mobi{\w}}					
\newcommand{\K}[1]{K_{#1}}					
\newcommand{\tK}{{\mathbf{K_A}}}					
\newcommand{\kr}[1]{\ensuremath{k_{r,#1}}}			
\newcommand{\mua}[1]{\ensuremath{\mu_{#1}}}			
\newcommand{\mun}{\mua\nw}					
\newcommand{\muw}{\mua\w}					
\newcommand{\qu}[1]{q_{#1}}					
\newcommand{\inRate}{\qu\coo}					
\newcommand{\vel}{\ensuremath{\mathbf{v}}}				
\newcommand{\pore}{\phi}					
\newcommand{\Sat}{S}						
\newcommand{\Sata}[1]{\Sat_{#1}}				
\newcommand{\eSa}[1]{\Sata{#1}^\ast}				
\newcommand{\sr}[1]{\ensuremath{\Sat_{\mathrm{res},#1}}}	
\newcommand{\pr}{\ensuremath{p}}				
\newcommand{\cp}{\ensuremath{C_p}}				

\newcommand{\mrcc}[4][\uu]{\ensuremath{{#1}^{#2}_{#3,#4}}}
\newcommand{\rv}{\theta}						
\newcommand{\vrv}{\ensuremath{\boldsymbol{\rv}}}		
\newcommand{\rvi}[1]{\rv_{#1}}					
\newcommand{\sdim}{\ensuremath{M}}				
\newcommand{\rf}{\Sat}						
\newcommand{\PrjSmb}{\ensuremath{\Pi}}
\newcommand{\Prj}[3][u]{\ensuremath{\PrjSmb^{\ifthenelse{\equal{#3}{-}}{{#2}}{{#3},{#2}}}\ifthenelse{\equal{#1}{-}}{}{\left[{#1}\right]}}}
\newcommand{\mean}{\mu_{S_g}}				
\newcommand{\std}{\sigma_{S_g}}				
\newcommand{\Response}{\Sat}				
\newcommand{\urate}{Q}					

\newcommand{\Nr}{\ensuremath{N_r}}				
\newcommand{\No}{\ensuremath{N_o}}				
\newcommand{\mi}[1]{\ensuremath{\mathfrak{#1}}}			
\newcommand{\smi}[2]{\ensuremath{\mi{#1}_{#2}}}			
\newcommand{\oP}{\ensuremath{P}} 				
\newcommand{\nP}{\ensuremath{\tilde{\oP}}} 			
\newcommand{\Sup}{\mathsf{I}}					
\newcommand{\elemA}[2][\Nr]{\ensuremath{\Sup^{#1}_{#2}}}	
\newcommand{\elem}[3][\Nr]{\ensuremath{\elemA[#1]{#2,#3}}}	
\newcommand{\cell}[1]{\ensuremath{E_{#1}}}			

\newcommand{\bpol}[1]{\ensuremath{\varphi_{#1}}}		
\newcommand{\Bpol}[1]{\ensuremath{\Phi_{#1}}}			
\newcommand{\BpolMR}[3][{\Nr}]{\ensuremath{\Bpol{#2,#3}^{#1}}}	

\newcommand{\multipoly}{\Phi}	
\newcommand{\NofCoeff}{N_P}	
\newcommand{\NofParam}{\sdim}	
\newcommand{\ExpansionOrder}{N_o}	

\newcommand{\sglevel}{\ell}	
\newcommand{\sgmultiindex}{\mi{i}}	
\newcommand{\sgmultilevel}{\mi{l}}	
\newcommand{\vli}{{\sgmultilevel, \sgmultiindex}} 
\newcommand{\Ind}{\mathcal{I}} 
\newcommand{\Indsg}{\mathcal{I}_\sglevel} 
\newcommand{\Indsgvl}{\mathcal{I}_\sgmultilevel}
\newcommand{\SGSpace}{V_\sglevel}         
\newcommand{\incrSpace}{W_\sgmultilevel}       
\newcommand{\asgm}{aSG$_\mathrm{m}$}
\newcommand{\asgb}{aSG$_\mathrm{b}$}
\newcommand{\asgms}{\asgm$\;$}
\newcommand{\asgbs}{\asgb$\;$}

\newcommand{\multihier}{\Phi_\vli}	
\newcommand{\unihier}{\phi^{(\sgmultilevel_j + 1)}}	
\newcommand{\sgcoeff}{v_\vli}	
\newcommand{\sgfunction}{\Response_{\Indsg}}
\newcommand{\sgfunctionwvli}{\Response_{\Indsg \setminus \{(\vli)\}}}

\newcommand{\kernel}{k} 
\newcommand{\xcenters}{\vrv} 
\newcommand{\Xset}{X} 
\newcommand{\kernelcoeff}{\boldsymbol{\alpha}} 
\newcommand{\kernelarg}{\vrv} 
\newcommand{\f}{f} 
\newcommand{\power}{P} 
\newcommand{\kernelmatrix}{K}
\newcommand{\RKHS}{\mathcal{H}} 
\newcommand{\kernelwidth}{\delta}
\newcommand{\bigN}{N}
\newcommand{\smallN}{n}

\newcommand{\vkoga}[2]{$\power$-VKOGA$_{#1}^{#2}$}

\newcommand{\outDim}{d}

\newcommand{\dataPoints}{\Theta}
\usepackage[table]{xcolor}
\newcommand{\good}{\cellcolor{green!15}Suitable}
\newcommand{\ok}{\cellcolor{yellow!15}Acceptable}
\newcommand{\bad}{\cellcolor{red!15}Not Suitable}
\newcommand{\goodFirstCol}{\cellcolor{green!15}Very common}
\newcommand{\okFirstCol}{\cellcolor{yellow!15}Common}
\newcommand{\badFirstCol}{\cellcolor{red!15}Uncommon}

\usepackage{color}

  \newcommand{\ab}[1]{{\color{black} #1}}
\newcommand{\comment}[1]{ }

\journalname{Computational Geosciences}
\begin{document}
\unitlength1cm

\title{Comparison of data-driven uncertainty quantification methods for a carbon dioxide storage benchmark scenario}

\titlerunning{Comparison of data-driven UQ methods for a carbon dioxide storage benchmark scenario}        

\author{Markus K\"oppel \and Fabian Franzelin \and Ilja Kr\"oker \and Sergey Oladyshkin \and Gabriele Santin \and Dominik Wittwar \and Andrea Barth \and Bernard Haasdonk \and Wolfgang Nowak \and Dirk Pfl\"uger \and Christian Rohde}


\institute{A. Barth \and B. Haasdonk \and M. K\"oppel* \and I.  Kr\"oker \and 
	   C. Rohde \and G. Santin \and D. Wittwar \\
	      IANS, Universtit\"at Stuttgart, Pfaffenwaldring 57, 
	      70569 Stuttgart, Germany \\
              \email{markus.koeppel@ians.uni-stuttgart.de} 
           \\ \\
            F. Franzelin \and D. Pfl\"uger \\
	      IPVS, Universtit\"at Stuttgart, Universit\"atsstra{\ss}e 38, 
	      70569 Stuttgart, Germany 
           \\ \\
           W. Nowak \and S. Oladyshkin \\
	      IWS, Universtit\"at Stuttgart, Pfaffenwaldring 5a, 
	      70569 Stuttgart, Germany \\ \\
}

\authorrunning{M.~K\"oppel {\itshape et al.}}

\date{November 2017}

\maketitle

\begin{abstract}
A variety of methods is available to quantify uncertainties arising with\-in the modeling of flow and transport in carbon dioxide storage, but there is a lack of thorough comparisons. Usually, raw data from such storage sites can hardly be described by theoretical statistical distributions since only very limited data is available.
Hence, exact information on distribution shapes for all uncertain parameters is very rare in realistic applications. We discuss and compare four different methods tested for data-driven uncertainty quantification based on a benchmark scenario of carbon dioxide storage. In the benchmark, for which we provide data and code, carbon dioxide is injected into a saline aquifer modeled by the nonlinear capillarity-free fractional flow formulation for two
incompressible fluid phases, namely carbon dioxide and brine. To cover different aspects of uncertainty quantification, we incorporate various sources of uncertainty such
as uncertainty of boundary conditions, of conceptual model definitions and of material properties. We consider recent versions of the following non-intrusive and
intrusive uncertainty quantification methods: arbitary polynomial chaos, spatially adaptive sparse grids, kernel-based greedy
interpolation and hybrid stochastic Galerkin. The performance of each approach is demonstrated assessing expectation value and standard deviation of the carbon dioxide saturation against a reference statistic based on Monte Carlo sampling. We compare the convergence of all methods reporting on accuracy with respect to the number of model runs and resolution. 
Finally we offer suggestions about the methods' advantages and disadvantages that can guide the modeler for uncertainty quantification in carbon dioxide storage and beyond.

\end{abstract}


\section{Introduction}
\label{intro}
Strong industrial development of the last century has led to a significant increase in public demand for different types of energy and, as a consequence, to an enormous increase
in demand for natural resources.
%
The subsurface is being used as storage plan for carbon dioxide (CO$_2$), nuclear waste or energy. In order to ensure efficient, safe and sustainable resource management, our society
needs a better understanding and improved predictive capabilities for subsurface problems. In particular, the ability to predict how the subsurface will react to planned
interventions is indispensable. However, subsurface flow and transport phenomena are complex and nonlinear. Moreover most subsurface systems are dominated by uncertainty where
external driving forces and material properties are observable only to a limited extent at high costs. Overall, this leads to an inherent uncertainty in all modeling endeavors and
in model-based predictions or decision support.

\subsection{Modeling carbon dioxide storage}
\label{intro_CO2}

Great research efforts have been directed towards understanding the processes of CO$_2$ storage in geological formation (GCS). It is currently being discussed intensively as an interim technology with high potential for mitigating CO$_2$ emissions (e.g. \cite{IPCC-Special}). GCS comprises capturing CO$_2$ at industrial facilities, compressing it into a fluid or supercritical state and disposing it in deep underground formations. The multiphase flow and transport processes involved
are strongly nonlinear. They include phase changes in the region of the critical point, effects such as gravity-induced fingering and convective mixing as well as geo-chemical and
geo-mechanical processes, etc. In order to describe the space-time evolution of injected CO$_2$ plumes and to investigate possible failure mechanisms of subsurface CO$_2$, (semi-)analytical solutions have been derived in \cite{Nordbotten2005}. A study that compares various simplifying semi-analytical models with complex numerical simulation tools was performed in \cite{Ebigbo2007}. The analysis in \cite{Birkholzer2009} focused on the effects of large-scale CO$_2$ leakage through low-permeability layers. Changes in pressure due to migration of fluids into the Above Zone Monitoring Interval of a geologic CO$_2$ site was studied in \cite{namhata2016probabilistic}. These
studies are cited here merely to provide a few examples. More detailed reviews are provided in, e.g., \cite{Bench,Ebigbo2007,IPCC-Special}. The current status
of CO$_2$ storage in deep saline aquifers with emphasis on modeling approaches and practical simulations is presented in \cite{celia2015status}. However, modeling
underground CO$_2$ storage involves uncertainty \cite{Hansson2009} due to the limited knowledge on subsurface properties (porosity, permeability, etc.), uncertainty in
physical conceptualization, uncertainty in boundary conditions and also human subjectivity in data interpretation \cite{OladNowak_AWR2010}. Thus, quantification of
uncertainty plays a key role in the development of CO$_2$ storage as a large-scale interim solution.

\subsection{Uncertainty quantification}
\label{intro_uq}



The main challenge in uncertainty quantification (UQ) is that brute-force stochastic simulation techniques (e.g. \cite{Helton2003}) are infeasible for large-scale
problems. Attempting to speed up uncertainty quantification can be subdivided into two principal ways: (1) developing analytical solutions, semi-analytical solutions,
conceptual simplifications, etc.; or (2) accelerating the forward modeling itself, e.g., using surrogate forward models such as response surfaces, emulators, meta-models,
reduced-order models, etc. The current paper focusses on the 2$^{\mathrm{nd}}$ way. 
A reasonably fast and attractive approach to quantify
uncertainty in CO$_2$ storage was pioneered in \cite{Olad_al_CG2009} via polynomial chaos expansion (PCE). This approach was further exploited during the last years.
However, there are other promising alternatives such as kernel methods and sparse grids that are discussed and employed in our current study.
The polynomial chaos expansion gained its popularity during the last decades due to an efficient massive reduction of computational costs in uncertainty quantification, see e.g.
\cite{foo_pcm_JCP2010,Ghanem_Spanos_1991_SFEM_book,Lin2009,zhang2016evaluation}. The key idea of PCE theory has been established by Wiener \cite{Wiener1938} and consists of projecting a full-complexity model onto orthogonal or orthonormal polynomial bases over the parameter space. Intrusive and non-intrusive approaches can be applied to estimate the involved projection integral in order to determine the form of the PCE. The non-intrusive approaches can be directly applied to the system of governing equations without any changes in simulation codes, however the intrusive approach demands rearranging of the governing equations.

Non-intrusive approaches like sparse quadrature \cite{Keese2003} and the probabilistic collocation method (\cite{Isukapalli1998,Li2007}) were applied to complex
and computationally demanding applications. 
PCE was combined with sparse integration rules
\cite{blatman2008sparse}, and an optimal sampling rule for PCE was proposed \cite{sinsbeckoptimal}. The adaptive multi-element
polynomial chaos approach \cite{MR2151997} was used to assure flexibility in treating the input distribution. A generalization of classical PCE was introduced in \cite{OladNowak_RESS2012} as arbitrary polynomial chaos (aPC) and provides a highly parsimonic and yet purely data-driven description of uncertainty. A recent extension to sparse approximation
via the moment-based aPC was presented in \cite{ahlfeld2016samba} and a multi-element aPC was introduced in \cite{alkhateeb2017data}. Additionally, a stochastic model calibration
framework was developed \cite{elsheikh2014efficient,OladClasNow_CG2013,Olad_EP2013} for CO$_2$ storage based on strict Bayesian principles combined with aPC.

Not only the approximation of models via various expansions, but also sampling of
the parameter space is a challenging procedure when the parameter space is high-dimensional. Sampling is directly
addressed via adaptive sparse grid techniques in the literature
\cite{Franzelin16Data,Jakeman11Characterization,Ma09adaptive}. Sparse
grids construct a potentially high-dimensional surrogate model using
Archimedes' hierarchical idea for quadrature. Each degree of freedom
adds the difference between the current approximation and the true
solution at the actual grid point to the approximation. In contrast to
global PCE techniques, for example, each degree of freedom has local
effect and the approximation does not suffer from the Gibbs phenomenon
even for basis functions of high polynomial order
\cite{Bungartz98Finite}. Moreover, highly efficient and parallel
implementations for the construction and the evaluation of the sparse
grid approximation are available
\cite{Pfander16New,Pflueger10Spatially}. Sparse grids are very
flexible and, hence, attractive to a large variety of applications
that arise in the context of uncertainty quantification: density
estimation \cite{Franzelin16Data,Peherstorfer13Model}, optimization
\cite{Valentin16Hierarchical}, etc.

As an alternative to polynomial or grid-based representation of the original physical model, other functions or kernels can be used. Kernel methods are well established techniques that found
broad application in applied mathematics \cite{Wendland2005} and machine learning \cite{SS02}. They are employed, e.g., for function
approximation, classification and regression. Since they are capable of working with meshless, i.e., scattered data in very high dimension, such methods are
particularly attractive in the
construction of surrogate models, where no restriction at all is imposed on the arbitrary location of the input data. In this context, greedy methods
\cite{DeMarchi2005,SchWen2000} have the additional advantage of providing sparse and thus fast-to-evaluate surrogate models \cite{SH2017a,Wirtz2015a},
while having provable error bounds and convergence rates \cite{SH16b,Wirtz2013}.

The most well-known intrusive approach is the stochastic Galerkin technique, which originated from structural mechanics \cite{Ghanem_Spanos_1991_SFEM_book} and has
been applied in studies for modeling uncertainties in flow problems (see e.g., \cite{Ghanem_Spanos_1991_SFEM_book,Matthies2005}). Several authors applied stochastic Galerkin methods to
hyperbolic problems. Apart from the hyperbolicity of the stochastic Galerkin system~\cite{MR2501693}, extensions to multi-element or
multi-wavelet based stochastic discretizations \cite{MR1199538,MR2151997} and also adaptivity for
the multi-wavelet discretization were provided ~\cite{tryoen2012}. The multi-element based hybrid stochastic Galerkin (HSG) discretization
used in this work and related stochastic adaptivity methods were introduced in~\cite{Barth201611,Buerger2012}. The application of HSG to two-phase
flow problems in two spatial dimensions was addressed in~\cite{knr2012} and extended to hyperbolic-elliptic systems in~\cite{Koeppel2017}.
In~\cite{Pettersson2016367} further improvements of intrusive stochastic Galerkin methods were suggested for the multi-wavelet discretization.


\subsection{Scope of the paper}
\label{intro_scope}

This work studies uncertainty quantification analysis for CO$_2$ storage using the modeling approaches discussed above. 
It seeks to offer a comparison that could be useful for further develepment considering uncertainty of boundary conditions, uncertainty of conceptual model definition and uncertainty of material properties.
Section \ref{CO2_problem} describes the physical model and Section \ref{sec:model} presents the case study
setup employed for the analysis. 
The key ideas of arbitrary polynomial chaos expansion, spatially adaptive sparse grids, kernel greedy interpolation and hybrid stochastic Galerkin are briefly described in Section \ref{sec:numex}, which also
demonstrates the performance of the introduced approaches against a reference solution. All mentioned methods have different nature and have their origins in different research areas.
However, we expect the identification of similarities in their performance.
Additionally, Section \ref{sec:discussion} presents the comparison between the methods in terms of
precision and corresponding computational effort.

We would like to invite the scientific community to participate and follow up on this work by comparing
and evaluating other available methods in this field based on the presented benchmark. Therefore,
we provide the corresponding input data and result files as well as the executables of the
deterministic code in~\cite{uqBenchmarkData}.


\section{Physical problem formulation}
\label{CO2_problem}

We consider a multiphase flow problem in porous media, where $\coo$ is injected into a deep aquifer and then spreads in a geological formation. This leads to a pressure
build-up and a plume evolution. In the current paper we consider a relatively simple model based on a benchmark problem defined by Class
et al. \cite{Bench} and reduce it considering the radial flow in the vicinity of the injection well to illustrate the performance of different methods for uncertainty quantification. The simplicity of the physical model is solely motivated by the high computational demand of our reference statistics based on Monte Carlo simulations, which we use for validation purposes. We assume that fluid properties such as density and viscosity are constant, all processes are isothermal, $\coo$ and brine are two separate and immiscible phases, mutual dissolution is neglected, the formation is isotropic rigid and chemically inert, and capillary
pressure is negligible.
In the following we describe the deterministic base model in more detail.

The initial conditions in the fully saturated domain include a hydrostatic pressure distribution which depends on the brine density.
The aquifer is initially filled with brine and CO$_2$ is injected at a constant rate at the center of the domain.
The lateral boundary conditions are constant Dirichlet conditions and equal to the initial conditions. All other boundaries are no-flow boundaries.
All relevant parameters used for the simulation are given in Table~\ref{tab:parameters}.

\begin{table}[!ht]
  \fontsize{8pt}{3ex}\selectfont
  \renewcommand{\arraystretch}{1.0}
  \begin{tabularx}{.45\textwidth}{ll}
    \toprule
    Parameter&Value\\
    \midrule
    CO$_2$ density,  $\varrho_g$&479 kg/m$^3$\\
    Brine density, $\varrho_w$&1045 kg/m$^3$\\
    CO$_2$ viscosity, $\mun$&3.950$\cdot10^{-5}$ Pa$\cdot$s\\
    Brine viscosity, $\muw$&2.535$\cdot10^{-4}$ Pa$\cdot$s\\
    Aquifer permeability, $\K{A}$&2$\cdot$10$^{-14}$ m$^2$\\
    Porosity, $\pore$&0.15\\
    Brine residual saturation, $\sr\w$ & 0.2\\
    CO$_2$ residual saturation, $\sr\nw$ & 0.05\\
    Injection well radius&0.15 m\\
    Injection rate, $\qu\coo$ &8.87 kg/s (1600m$^3$/d)\\
    Dimension of model domain, $\rd_{\max}$& $500$ m \\
    Simulation time, $t$&100 days\\

    Saturation on the left boundary & $0.8$  \\
    Injection pressure $\pr_{\max}$& 320 bar \\
    Pressure right boundary $\pr_{\min}$ & 300 bar\\
    Mean mobility value $\mobi{}$ & $1.0\cdot10^{4}$ (Pa$\cdot$s)$^{-1}$\\
    \bottomrule
  \end{tabularx}
  \caption{Simulation parameters.}
  \label{tab:parameters}
\end{table}

For time $T > 0$, domain $\setD\subset\setR^3$ and $(\vx,t)\in\setD_T:=\setD\times(0,T)$, the well-known two-phase flow
equations obtained from mass balances of both fluid phases and the multiphase version of Darcy's law can be
reformulated by means of the fractional flow formulation \cite{Nordbotten2011} given by the following
system of equations
\begin{eqnarray}
\displaystyle  \label{2phase_fracflow}
\pore \frac{\partial \Sata\alpha}{\partial t} + \nabla \cdot (\vel \fxa\alpha) - \qu\alpha = 0, & & \quad\mbox{ in }\setD_T,  \\
\displaystyle  \label{2phase_darcy_total}
\vel = -  \mobi{} \tK \nabla \pr, & & \quad\mbox{ in }\setD_T,	\\
\displaystyle  \label{2phase_massb_total}
\nabla \cdot \vel = \qu\w+\qu\nw, & & \quad\mbox{ in }\setD_T,	\\
\displaystyle  \label{2phase_ic}
\Sata\alpha(\cdot,0) = \Sata{0,\alpha}, & &\quad\mbox{ in }\setD,
\end{eqnarray}
where the subscript $\alpha \in \{\w,\,\nw\}$ stands for the brine (water) phase ($\alpha=\w)$ and the $\coo$-rich (gas) phase ($\alpha=\nw)$, respectively, and the absolute permeability $\tK$, porosity $\pore$ and the sources/sinks $\qu\alpha$ are given parameters. Combined with the contraint $\Sata\w + \Sata\nw = 1$, the primary variables of the system \eqref{2phase_fracflow}-\eqref{2phase_massb_total} are the phase saturation $S_\alpha$, the total velocity $\vel$ and the global pressure $\pr$. The fractional flow function $\fxa\alpha:=\mobi\alpha/\mobi{}$ and the mean mobility function  $\mobi{}:=\mw+\mn$ are nonlinear functions of the saturation $S_\alpha$. Both are defined via the mobilities $\mobi\alpha:=\kr\alpha(\eSa\alpha) / \mua\alpha$, $\alpha=\w,\,\nw$, with dynamic viscosities  $\mua\alpha$ and the relative permeabilities  $\kr\w$ and $\kr\nw$ given by
\begin{eqnarray}
\label{rel_perm}
\kr\nw(\eSa\nw)&:=&\left(\eSa\nw(\Sata\nw)\right)^2, \\
\kr\w(\eSa\w)&:=&\left(1-\eSa\w(\Sata\w)\right)^2.
\end{eqnarray}
Moreover, $\eSa\alpha = (\Sata\alpha-\sr\alpha)/(1-\sr\alpha)$ is the effective saturation, where $\sr\alpha$ denotes the residual saturations of the fluid phases. Insertion of \eqref{2phase_darcy_total} in \eqref{2phase_massb_total} yields
\begin{eqnarray}
\label{2phase_sat_sum}
\displaystyle
\nabla \cdot \left(\mobi{} \tK \nabla \pr \right) = \qu\w+\qu\nw.
\end{eqnarray}

\subsection{Radial flow equations}

\comment{Internal remark: $grad \fx = d\fx/d\rd$ and $div \fx = 1/\rd d(\rd\fx)/d\rd$}
\comment{Valid iff $f$  radial symmetric.}

We consider radial flow in the vicinity of the injection well
in the homogeneous reservoir with scalar absolute permeability $\K{A}$.
Hence, the governing equation for pressure
(\ref{2phase_sat_sum})  can be written in the following form
\begin{eqnarray}
\label{press_1}
\frac{1}{\rd} \frac{\partial}{\partial \rd}  \left(-\rd \mobi{}\K{A} \frac{\partial\pr}{\partial \rd} \right) &= \qu\w + \qu\nw,
\end{eqnarray}
where $r$ is the radial coordinate and $\qu\w+\qu\nw$ controls the injection rate in the well.
%
Since only $\coo$ is injected, i.e. $\qu\w+\qu\nw=\qu\coo$, equation (\ref{press_1}) can be integrated as
\begin{eqnarray}
\label{press_2}
-\rd \lambda\K{A} \frac{\partial\pr}{\partial \rd} = \qu\coo\cp,
\end{eqnarray}
with constant $\cp$.
The solution of equation (\ref{press_1}) can be written in the closed analytical form
\begin{eqnarray}
\label{press_anal}
\pr(\rd)=\pr_{\max}- \frac{\qu\coo\cp}{\mobi{}\K{A}} \ln\rd, \quad \rd\in[1,\rd_{\max}],
\end{eqnarray}
with injection pressure $\pr_{\max}$ and $\cp$ given by 
\[
\cp:=\frac{\pr_{\max} - \pr_{\min}}{\inRate \ln  \rd_{\max}} \K{A}\mobi{}.
\]
Using the parameters in Table~\ref{tab:parameters}, 
we get $\cp=3.48\cdot10^{-3}$.
\comment{We control the boundary condition over the injection rate.

The equation \ref{press_anal} show logarithmic distribution of the pressure that is typical one. We can add here some 1D or
2D radial plot,just for illustration.

Hence, substituting the equation \ref{press_anal} into the equation for the velocity $V=-\lambda\tK\nabla\pr$ we will obtain that $V=Q_\w + Q_\nw$. So $V$ is only constant and no
relation between $\tK$ and $\phi$ needed. Do not worry we will introduce a bit of nonlinear once we reformulate saturation equation in the radial coordinate system}
%
We reformulate  equation (\ref{2phase_fracflow}) for the gas phase using the radial coordinate system and \eqref{press_2} to obtain
\begin{eqnarray}
\label{2phase_fracflow_rad_2}
\displaystyle
\pore \frac{\partial \Sata\nw}{\partial t} - \frac{1}{\rd} \frac{\partial}{\partial \rd}  \left(\inRate\cp \fxg \right) - \inRate = 0.
\end{eqnarray}
Because the velocity is constant, equation \eqref{2phase_fracflow_rad_2} does not depend
on the absolute permeability as the porous medium is assumed to be homogeneous.

\subsection{Hyperbolic solver}\label{sec:hyp_solver}

In order to discretize the hyperbolic transport equation \eqref{2phase_fracflow_rad_2} in the physical space,
we apply a semi-discrete central-upwind finite volume scheme introduced in \cite{NUM:NUM20049}. 
Central-upwind schemes are typically characterized by robustness and high accuracy up to second order.
In contrast to, e.g., Godunov-type solvers \cite{leveque1992}, where analytical knowledge about the front propagation is essential, central-upwind schemes only require information about propagation speeds. By construction,
the artifical viscosity inherent to the scheme is adapted to the discrete solution and thus leads to lower
numerical dissipation compared to other schemes such as Lax-Friedrichs \cite{leveque1992}. For the temporal discretization the Runge-Kutta method of second order is applied.

Let $\rdi{j}=j\Delta\rd, \; \rdi{j\pm1/2}=(j\pm1/2)\Delta\rd$, $j=1,\ldots,N_e$, with the number of elements $N_e$ and $\cell{j+\half}=(\rdi{j},\rdi{j+1})$,
where $\Delta\rd=1$ represents the uniform, radial mesh size in the physical space.
For the sake of brevity we will denote the unknown by $\Sat := \eSa\nw$ and $\urate:=\inRate\cp$.
Then the semi-discrete scheme reads
\begin{equation}\label{eq:fvscheme}
 \pore\frac{d}{dt}\bar{\Sat}_{j+\half}(t) := - \frac{\urate}{\rd}\, \frac{\nfx{j+1}(t)-\nfx{j}(t)}{\Delta\rd} + \inRate,
\end{equation}
and the numerical flux function $\nfx{j}(t)$ is given by
\begin{equation*}
 \nfx{j}(t) := \frac{a_j^+\fx(\Sat_j^-) - a_j^-\fx(\Sat_j^+)}{a_j^+ - a_j^-}  +
		  \frac{a_j^+ a_j^-}{a_j^+ - a_j^-} (\Sat_j^+ - \Sat_j^-),
\end{equation*}
with the cell averages $\bar{\Sat}_{j+\half}(t)$, the right- and left-sided local speeds $a_j^\pm$, and the
piecewise linear reconstructions $\Sat_j^\pm$ at the interface points $\{\rdi{j}\}$. The initial values of the
cell averages can be computed by
 $\bar{\Sat}_{j+\half}^0 = \frac{1}{\Delta\rd} \int_{\cell{j+\half}} \Sat_0(\x)d\x$.
Note that the CFL condition and the local speeds depend on the derivative of the flux function $\fx$.
For a more detailed description of the used finite volume scheme we refer to \cite{NUM:NUM20049}. 

\section{Benchmark case study setup: modeling parameters and quantity of interest}\label{sec:model}

\comment{The usage of the fractional flow form instead of the standard multi-phase flow form is briefly discussed in
\cite{Nordbotten2011} and simplifies numerical implementations. Moreover it justifies the numerical treatment
of the transport equation instead of an analytical representation, in particular if the $S^{res}_n\neq 0$.
However this is at the expense of a detailed comparability to \cite{Court2012134,Nordbotten2011}.
Thus we can use it rather as a motivation to use 1D in the deterministic space. 
}

In our benchmark case, we analyze the joint effect of various sources of uncertainty. Typically, the following types of uncertainty can occur during the reservoir screening stage:
uncertainty of boundary conditions, uncertainty of conceptual model definition and uncertainty of material properties.

We consider the uncertainty of boundary conditions via the injection rate $\inRate$. The reservoir pressure can thus be seen as a function of the injection rate
$\inRate$,
\begin{eqnarray}
\label{u_injection_rate}
\displaystyle
\pr(\rd,\rv_1)&=\pr_{\max}-\frac{\urate(\rv_1)}{\mobi{}\K{A}}\ln \rd, \quad r\in[1,\rd_{\max}]\,,
\end{eqnarray}
where $\urate(\rv_1):=\cp\inRate\left(1+\rv_1 \right)$ and $\rv_1$ denotes the random variable.
Conceptual model uncertainty is introduced via uncertainty in the relative permeability definitions $\kr\nw$ and $\kr\w$ (see \cite{Court2012134}) which we extend to
%
\begin{align}
\label{u_relative_permeability}
\kr\nw(\eSa\nw,\rv_2)&:= {(\eSa\nw)}^{\rv_2},\\
\kr\w(\eSa\w,\rv_2)&:= \left(1-\eSa\w\right)^{\rv_2},
\end{align}
with random variable $\rv_2$.
Generally, variations of the relative permeability degree have a strong impact on the fractional flow function.
Uncertainty of material properties are represented via uncertainty of reservoir porosity. In the current study, we have aligned the distribution of the reservoir porosity with data from the U.S.
National Petroleum Council Public Database (see also \cite{Kopp2009}). Thus, the reservoir porosity can be written in the form $\pore(\rv_3):=\rv_3$ with random variable $\rv_3$.

The uncertain parameters represent the input parameters of equation (\ref{2phase_fracflow_rad_2}) and can be written as an
$\sdim$-dimensional random vector $\vrv := \{\rvi{1},\ldots,\rvi{\sdim}\}$, $\sdim\in \setN$  with $\sdim=3$ for the current case study.
We will assume that each random variable
$\rvi{i}$ ($i\,=\,1,\ldots,\sdim$) is independent and $\vrv\in\Lp[2](\WR)$ on the probability space $\WRM$, where $\WR$ is a sample space with a $\sigma$-algebra $\Field$ and probability measure $\Pm$. The distributions are chosen to reflect the situation of site screening, where
site-specific data and data that allow detailed description of injection strategy, fluid properties and geology are not yet available. In this stage, one has to resort to databases and expert elicitation that represent properties of supposedly similar sites as prior knowledge.

\begin{figure*}[htb]
\centering
\subfigure{\includegraphics[width=.32\textwidth]{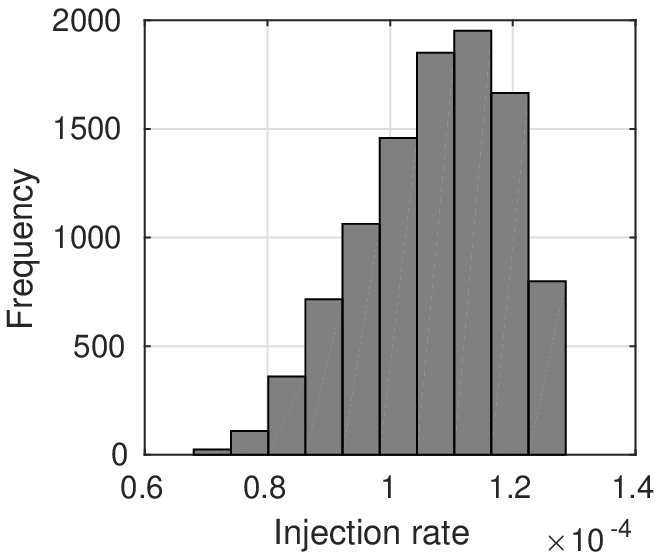}}
\subfigure{\includegraphics[width=.31\textwidth]{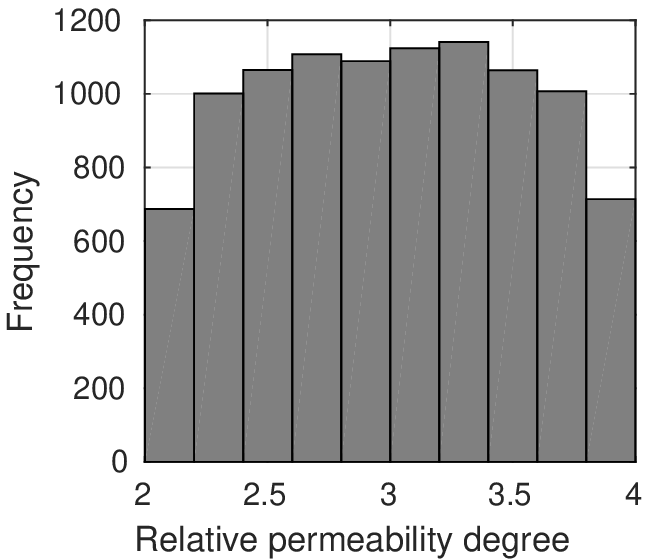}}
\subfigure{\includegraphics[width=.32\textwidth]{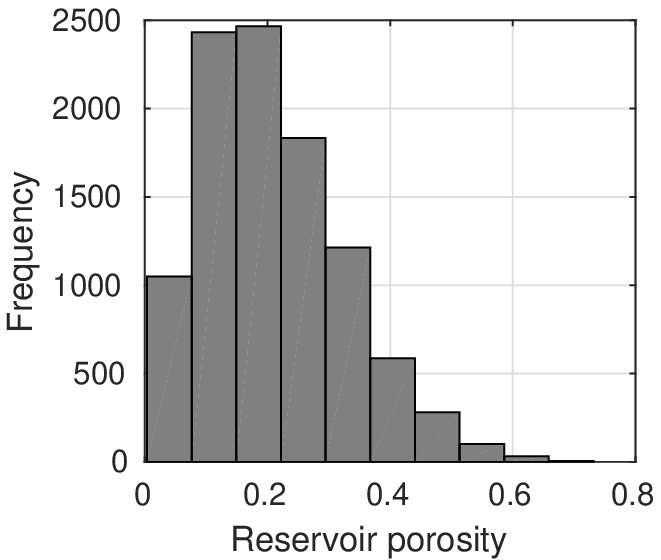}}
\caption{Parameter distributions of injection rate [m$^3$/s], relative permeability degree [-] and reservoir porosity [-].}
\label{fig:input-data-distribution}
\end{figure*}

From the random vector $\vrv$, we generate a set of 10.000  samples denoted $\dataPoints\subset \setR^3$  to construct an exact 
reference solution for the moments of the quantity of interest via the statistics, and to construct a data-driven framework for the methods of consideration.
Fig.~\ref{fig:input-data-distribution} shows univariate histograms of $\dataPoints$.
We stress that the data set $\dataPoints$ is deployed by all methods without prior knowledge on the distribution.

%
%

\begin{figure*}[htb]
\centering
\subfigure{\includegraphics[width=.48\textwidth]{\picdir/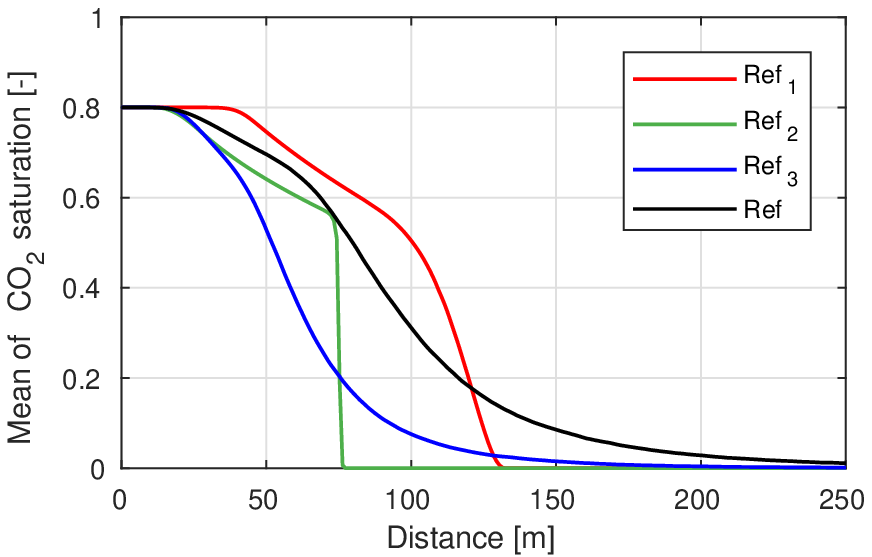}}\hfill
\subfigure{\includegraphics[width=.48\textwidth]{\picdir/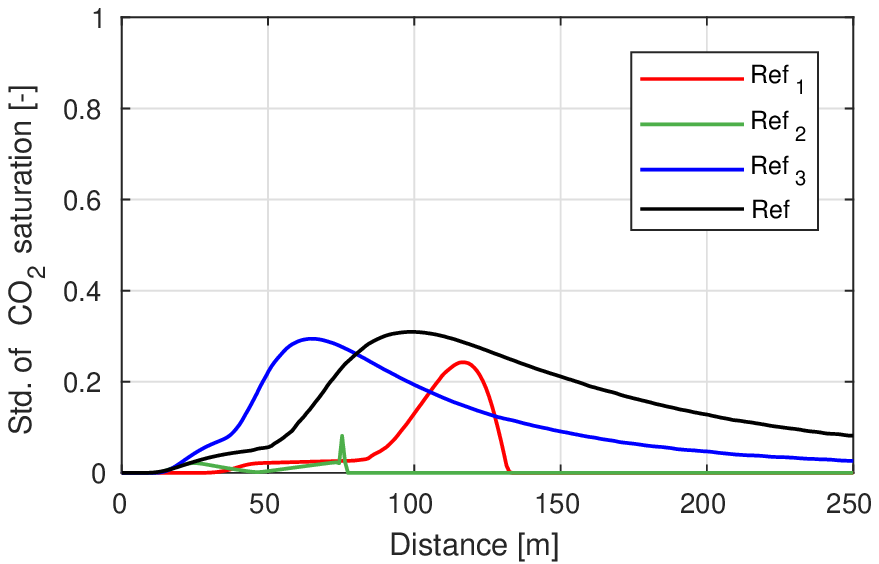}}
\caption{\label{fig:FktX} Mean and standard deviation of CO$_2$ saturation at $t=100$~days based on Monte Carlo sampling using $10^4$ samples: Ref$_{\,1}$ -
uncertain injection rate; Ref$_{\,2}$ - uncertain relative permeability degree; Ref$_{\,3}$ - uncertain porosity;
Ref - joint uncertainty.
}
\end{figure*}

This study
quantifies stochastic characteristics of the flow using mean value $\mean$ and standard deviation $\std$ of $\coo$ saturation as a function of space and time. 
Fig.~\ref{fig:FktX} shows the statistical reference solution for the mean value and standard deviation of the $\coo$ saturation after $100$ days based on the set of samples $\dataPoints$.
Apart from  the global influence of the uncertain parameters onto the output statistics, Fig.~\ref{fig:FktX} also illustrates the individual impact of each analyzed parameter.
One can observe that the uncertainty in the degree of the relative permeability does not influence the dynamics of the saturation significantly. The injection rate and the
reservoir porosity are the main cause of uncertainty in the $\coo$ saturation.


%

\section{Uncertainty quantification methods}\label{sec:numex}
In this section we briefly introduce four different methods for uncertainty quantification and discuss some of their properties. We compare them with the data-based statistics generated by Monte Carlo sampling for mean value and standard deviation of the quantity of interest (black line in Fig.~\ref{fig:FktX}).

\subsection{Non-intrusive arbitrary polynomial chaos expansion}
\label{sec:nipc}

We briefly introduce arbitrary polynomial chaos (aPC) techniques that are employed to construct a global response surface which captures the dependence of the model on the data set. We consider space and time dependent model response 
$\Sat(\rd, t;\vrv)$ of the $\coo$ saturation.  
According to Wiener \cite{Wiener1938} the dependence of the model output on all input parameters is expressed via projection onto a multi-variate polynomial basis (see e.g. \cite{Ghanem_Spanos_1991_SFEM_book}), such that the model output $\Sat$ can be approximated by the polynomial chaos expansion
%
\begin{eqnarray}
\label{PCE}
\Sat(\rd, t;\vrv) \approx \sum_{i=0}^{\NofCoeff} \Sat_{i}(\rd,t) \multipoly_i(\vrv),
\end{eqnarray}
where $\NofCoeff$ is the number of multi-variate polynomial basis functions $\multipoly_i(\vrv)$ (see e.g. \cite{Ghanem_Spanos_1991_SFEM_book}) and corresponding 
coefficients $\Sat_i(\rd,t)$. It depends on the total number of input parameters $\NofParam$ and on the order $\ExpansionOrder$ of the polynomial representation: 
$\NofCoeff = (\NofParam+\ExpansionOrder)!/(\NofParam!\ExpansionOrder!)-1$. The coefficients $\Sat_{i}(\rd,t)$ in equation (\ref{PCE}) quantify the dependence of the model 
response $\Sat(\rd, t; \vrv)$ on the input parameters for each desired point in space $\rd$ and time $t$.

%
%
%

We follow a recent generalization of the polynomial chaos expansion known as the arbitrary polynomial chaos (aPC). The aPC technique adapts to arbitrary probability distribution shapes of the input parameters and can be inferred from limited data through a few statistical moments \cite{OladNowak_RESS2012}. The necessity to adapt to
arbitrary distributions in practical tasks is discussed in more detail in \cite{OladNowak_AWR2010}. Thus, we explore a highly parsimonic and purely data-driven description of
uncertainty via aPC and directly incorporate the available data set of size $10^4$ illustrated in Fig.~\ref{fig:input-data-distribution} without any use of exact forms
of probability
density functions. For that, we compute $2N_o$ raw statistical moments from $10^4$ realisations and then we construct the orthonormal polynomial basis of order $N_o$
according to the matrix equation introduced in  \cite{OladNowak_RESS2012}.  Note that the orthonormal basis can be also obtained via recursive relations (see Chapter 22 of \cite{Abramowitz1965}), via Gram-Schmidt orthogonalization (see \cite{Witteveen2007}) or via the Stieltjes
procedure \cite{Stieltjes1884}.

The polynomial representation in equation (\ref{PCE}) is fully defined via the unknown expansion coefficients $\Sat_{i}(\rd,t)$. These coefficients can be determined using
intrusive or non-intrusive approaches. The intrusive approach requires manipulation of the governing equations and will be discussed in Section \ref{sec:hsg} via hybrid stochastic
Galerkin. In the current Section \ref{sec:nipc} we follow the non-intrusive way where no modifications are required for the system of governing equations 

\begin{figure*}[htb]
\subfigure{\includegraphics[width=.48\textwidth]{\picdir/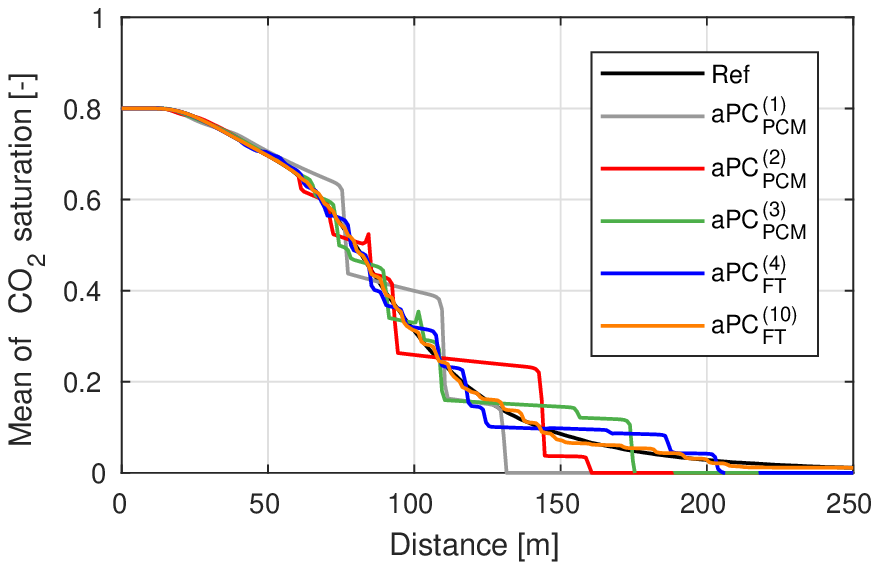}}\hfill
\subfigure{\includegraphics[width=.48\textwidth]{\picdir/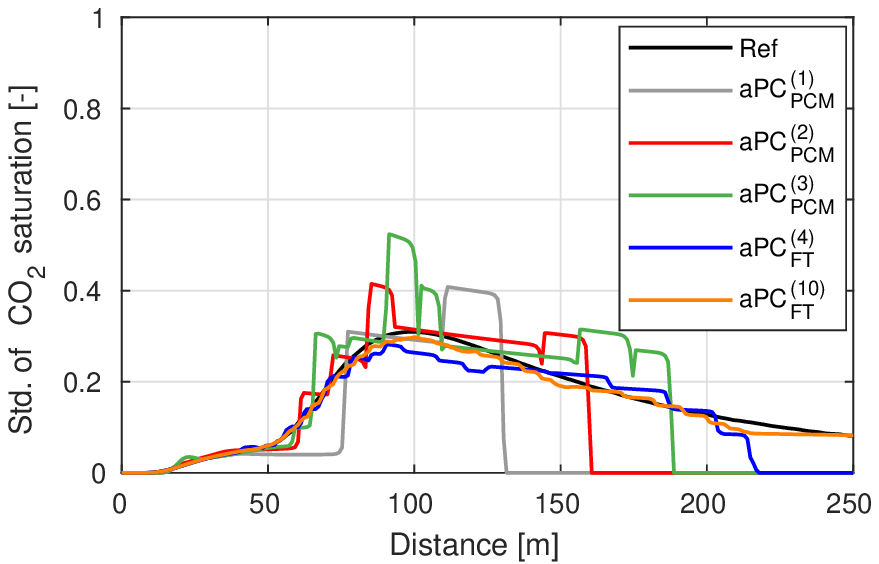}}
\caption{\label{fig:aPC_full} Mean and standard deviation of CO$_2$ saturation estimated via aPC expansion with probabilistic (PCM) and least-squares collocation method (FT):
 aPC$^{(1)}_{PCM}$ - $1^{st}$ order expansion; 
 aPC$^{(2)}_{PCM}$ - $2^{nd}$ order expansion; 
 aPC$^{(3)}_{PCM}$ - $3^{rd}$ order expansion; 
 aPC$^{(4)}_{FT}$ - $4^{th}$ order expansion; 
 aPC$^{(10)}_{FT}$ - $10^{th}$ order expansion; 
Numbers in $(\cdot)$ indicate the number of model runs.}
\end{figure*}

As the computationally cheapest version we apply the non-intrusive probabilistic collocation method (PCM) \cite{Li2007,Olad_al_CG2009}. The method is based on a minimal chosen set of model evaluations, each with a defined set of model parameters (called collocation points) that is related to the roots of the polynomial basis via optimal integration theory \cite{Villadsen1978}. Fig.~\ref{fig:aPC_full} shows mean and standard deviation of the CO$_2$ saturation estimated via aPC expansion based on the
probabilistic collocation method and also shows the \ab{statistical} reference solution. As expected, the strong discontinuity of the original physical model introduced due to the CO$_2$ displacement front poses challenges for the global polynomial representation. Nevertheless, the estimation of the mean value is acceptable. However, increasing the expansion order does not necessary lead to improvement of the results, especially for the variance estimation. Hence, a moderate expansion order can be seen as adequate compromise between accuracy and computational efforts.
Additionally, the expansion order is only justified if accompanied by reliable statistical information, because incomplete statistical information limits the utility of polynomial chaos expansions \cite{oladyshkin2018incomplete}.

As a computationally very demanding alternative to the probabilistic collocation method, we also employ the least-squares collocation method (e.g. \cite{Moritz1978}) for
constructing the expansion coefficients on a full tensor grid of collocation points. Fig.~\ref{fig:aPC_full} also shows mean and standard deviation of CO$_2$ saturation
estimated via aPC expansion based on the least-squares collocation method against the \ab{statistical} reference solution. The least-squares collocation method based on the full tensor (FT)
grid helps to overcome the typical oscillation problem of polynomials for high order expansions. However, due to the curse of dimensionality for tensor grids,
this approach has an extremely high computational effort if more than a single parameter is of interest. 

\label{sec:nipc_results}

%

\subsection{Spatially adaptive sparse grids}\label{sec:sgrid}

In this section we introduce regular sparse grids according to
\cite{Zenger91Sparse}. We follow the approach of higher-order basis
functions that have been presented in \cite{Bungartz98Finite} and
extended in \cite{Pflueger10Spatially} with proper extrapolation
schemes. Furthermore, we present the concept of spatially adaptive
refinement following \cite{Pflueger12Spatially} and provide refinement
criteria in the context of data-driven uncertainty quantification
\cite{Franzelin15Non,Ma09adaptive}.

Let
$\sgmultilevel := \{\sgmultilevel_1, \dots, \sgmultilevel_\NofParam\}$
be a multi-index with $\sgmultilevel_j > 0$ and dimensionality
$0 < \NofParam$. We define a level-index set for some $\sgmultilevel$
as
\begin{equation}
  \label{eq:sparse-grid-level-index-sets}
  \Indsgvl := \{\sgmultiindex \in \setN^\NofParam \colon 1\leq \sgmultiindex_j<2^{\sgmultilevel_j}, \sgmultiindex_j \text{ odd}, j=1,\dots,\NofParam\},
\end{equation}
that defines grid points located at
$\rv_{\sgmultilevel_j, \sgmultiindex_j} := 2^{-\sgmultilevel_j}
\sgmultiindex_j$. The multivariate basis functions are centered at the
grid points and are defined as the tensor product of one-dimensional,
local polynomials
\begin{equation}
  \label{eq:sg-tensor-product}
  \multihier(\vrv) :=
  \begin{cases}
    \prod_{j=1}^\NofParam \unihier_{\sgmultilevel_j,
      \sgmultiindex_j}(\rv_j) & \text{for } \vrv \in X_\vli \\
    0 & \text{else} \;,
  \end{cases}
\end{equation}
where
$X_\vli := \bigtimes_{j=1,\dots,\NofParam} [2^{-\sgmultilevel_j}
(\sgmultiindex_j - 1), 2^{-\sgmultilevel_j} (\sgmultiindex_j + 1)]$
and $\sgmultilevel_j + 1$ is the polynomial degree in direction $j$.

Sparse grids use these functions to form a hierarchical basis in order
to overcome the curse of dimensionality to some extent. The
level-index sets $\Indsgvl$ define a unique set of hierarchical
increment spaces
$\incrSpace := \textrm{span}\{\multihier(\vrv) \colon \sgmultiindex
\in \Indsgvl\}$ that add the difference between the approximation on
smaller levels (componentwise) and the actual level.
Due to this hierarchical character, one can sort the increment spaces
according to their benefit to the approximation of functions from
various function spaces and leave out the less important ones. If the
contribution of each $\incrSpace$ to the approximation is measured
with respect to the $L^2$-norm, then the optimal sparse grid
level-index set is obtained as
\begin{equation}
  \label{eq:sparse-grid-index-set}
  \Indsg :=
  \bigcup_{\sgmultilevel \in \setN^\NofParam \colon |\sgmultilevel|_1 \leq \sglevel + \NofParam
    - 1} \{(\sgmultilevel, \sgmultiindex) \colon \sgmultiindex \in
  \Indsgvl \}\;, |\sgmultilevel|_1 := \sum_{j=1}^\NofParam \sgmultilevel_j\;,
\end{equation}
with $\sglevel \in \mathbb{N}$ being the regular level of the grid. A
regular sparse grid function
$\sgfunction \in \SGSpace := \bigoplus_{|\sgmultilevel|_1 \leq
  \sglevel + \NofParam - 1} \incrSpace$ that approximates the model
output $\Response$ is written as
\begin{equation}
  \label{eq:regular-sparse-grid-function}
  \Response(\rd, t;\vrv) \approx \sgfunction(\rd, t; \vrv) := \sum_{(\vli) \in
    \Indsg} \sgcoeff(\rd, t) \multihier(\vrv)\;,
\end{equation}
where the $\sgcoeff(\rd, t) \in \setR$ are called hierarchical
coefficients.

The number of grid points $|\Indsg|$ is significantly reduced compared
to a full grid with the same spatial resolution in each direction. At
the same time the interpolation error of a sparse grid differs just by
a logarithmic factor compared to a full grid and is, hence, only
slightly worse \cite{Bungartz98Finite,Zenger91Sparse}.

One can interpret such a regular sparse grid as the result of an
a-priori adaptivity. Spatially adaptive sparse grids add a second
level of refinement: Grid points are added iteratively where the local
error of the approximation is largest. Refinement criteria estimate
these local errors with respect to some target quantity. In this paper
we use a weighted $L^2$-refinement method and enforce balancing
\cite{Bungartz03Multivariate}. It defines a ranking for all
level-index pairs as
\begin{equation}
  \label{eq:adaptivity-l2-norm-result}
  \max_{(\vli) \in \Ind} \|\Response -
  \sgfunctionwvli\|_{L^2(\Omega)} \approx \max_{(\vli) \in \Ind} |v_\vli| \|\multihier\|_{L^2(\Omega)}\;,
\end{equation}
where $\Ind$ is an adaptive sparse grid index set. To refine, we add all the
hierarchical successors of $(\vli) \in \Ind$,
\begin{equation}
  \label{eq:hierarchical-successors}
  \begin{aligned}
    \{&((\dots, \sgmultilevel_m + 1, \dots), (\dots, 2\sgmultiindex_m + 1, \dots)), \\
    & ((\dots, \sgmultilevel_m + 1, \dots), (\dots, 2\sgmultiindex_m - 1, \dots))\}_{m=1}^\NofParam\;,
  \end{aligned}
\end{equation}
that are not yet part of $\Ind$ starting with the one that has the
largest rank.

To describe the uncertainty, we use a sparse grid probability density
function \cite{Franzelin16Data} based on the input data to approximate
$\|\multihier\|_{L^2(\Omega)}$. This way, we can start with a purely
data-driven description and arbitrary densities without any need for
derived analytical forms or independence of the respective probability
functions.

%

\begin{figure*}[htb]
\subfigure{\includegraphics[width=.48\textwidth]{\picdir/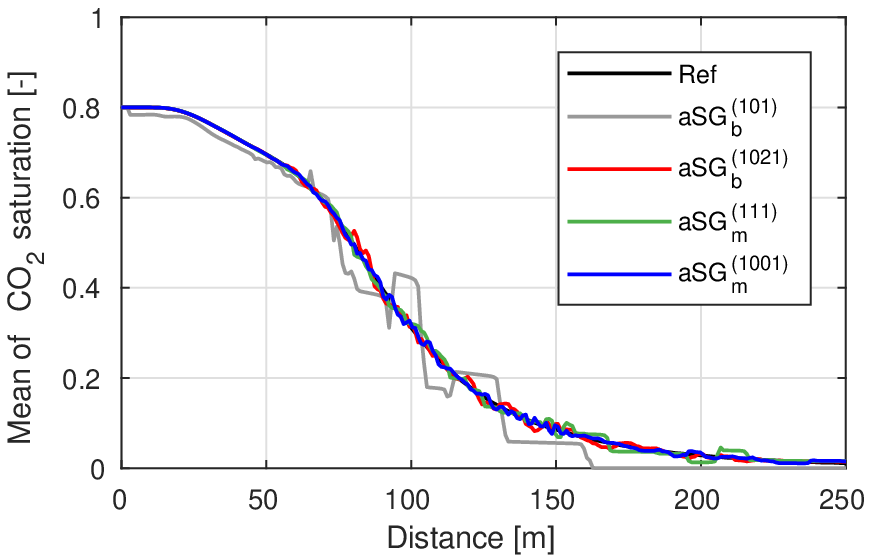}}\hfill
\subfigure{\includegraphics[width=.48\textwidth]{\picdir/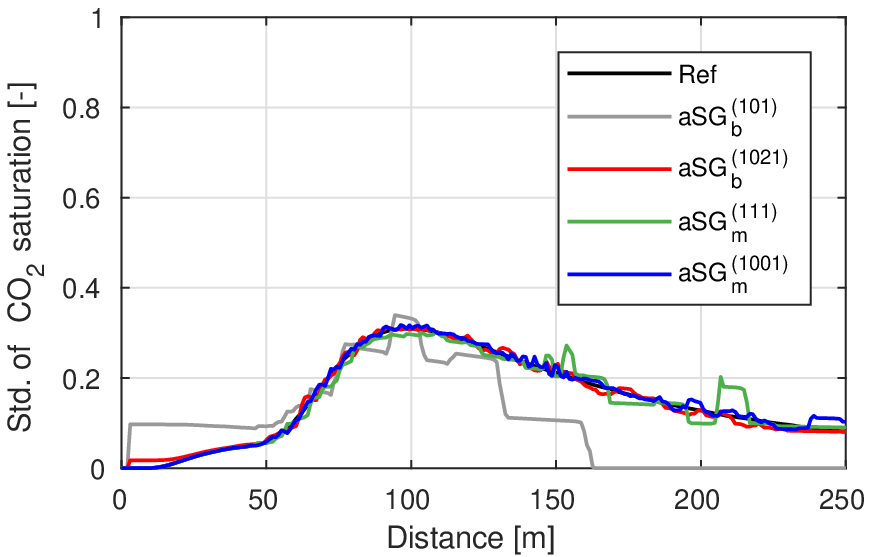}}
\caption{\label{ff:fig:asg_l2_full} %
  Mean and standard deviation of CO$_2$ saturation estimated via
  adaptive sparse grids with boundary points (\asgb) and modified basis (\asgm):
  aSG$^{(101)}_{\textrm{b}}$ - after $3$ refinement steps;
  aSG$^{(1021)}_{\textrm{b}}$ - after last refinement iteration;
  aSG$^{(111)}_{\textrm{m}}$ - after $5$ refinement steps;
  aSG$^{(1001)}_{\textrm{m}}$ - after last refinement iteration;
  Numbers in $(\cdot)$ indicate the number of model runs / grid points. }
\end{figure*}

For this model problem we distinguish two types of sparse grids: The
first model spends grid points directly at the boundary
\cite[p.15]{Pflueger10Spatially} to which we refer to as \asgb. As a
second model, we consider modified piecewise polynomial basis
functions \cite[p.24]{Pflueger10Spatially} with linear extrapolation,
which we write as \asgm. Both sparse grid surrogates are constructed
as follows:
\begin{description}
\item[Step 1:] Start with a regular sparse grid $\Ind$ of
  $\sglevel=1$.
\item[Step 2:] Compute the ranking
  \eqref{eq:adaptivity-l2-norm-result} of all sparse grid points for
  which not all successors \eqref{eq:hierarchical-successors}
  exist. Add the successors of the two highest ranked level-index
  pairs to $\Ind$.
\item[Step 3:] Make sure that all hierarchical ancestors of each grid
  point exist and that each grid point has either none or two
  hierarchical successors in each direction.
\item[Step 4:] Run the model problem at the new grid points and
  construct the new interpolant.
\item[Step 5:] Continue with step 2 until a maximum number of grid
  points is reached.
\end{description}
Sample results for the expectation value and the variance of the model
problem are shown in Fig.~\ref{ff:fig:asg_l2_full}. Both estimated
quantities using a sparse grid with, for example, $101$ boundary
points differ significantly from the \ab{statistical} reference value.
Most of the grid points are located at the boundary of the domain,
which makes this approach unfeasible for problems with small
computational budget. This problem can be solved with \asgms. Hence,
we observe a better approximation of the expectation value and the
variance already for smaller grids. Nevertheless, the accuracy of the
moments with respect to the reference solution increase with
increasing grid size in both cases.

\subsection{Kernel greedy interpolation}\label{sec:rb}
To start, we emphasize that we work here  with the discretization of the saturation provided by the hyperbolic solver, thus we understand
$\rf(\rd,t,\vrv)$ as a
function $\Xset\to \setR^{d}$, $\Xset\subset\setR^3$, $d:=250$, mapping the uncertain parameters to the spatial discretization consisting of 250 cells at final time.
Kernel-based approximation methods construct a surrogate $\rf_N(\vrv)$ of $\rf(\rd,t,\vrv)$ based on a set $\Xset_{\bigN}\subset\Xset$ of $\bigN\in\setN$ input
parameters and the corresponding output computed by the solver.
The surrogate model can then be rapidly evaluated on the large set $\dataPoints\subset\Xset$ of input parameters of the reference solution, and the mean and
variance are calculated as the mean and variance of these evaluations. 

We give only a brief overview of kernel methods, and refer to \cite{Wendland2005} for a detailed treatment.
The saturation  is approximated as a vector-valued linear combination
\begin{equation}\label{eq:kernel expansion}
 \rf(\rd,t,\vrv)\approx \rf_N(\vrv):= \sum\limits_{i=1}^{\bigN} \kernel(\kernelarg,\xcenters_i)\kernelcoeff_i   ,
\quad \kernelarg\in \Xset,
\end{equation}
where $\kernel : \Xset \times \Xset \rightarrow \setR$ is a symmetric kernel function, $\xcenters_i \in \Xset_{\bigN} \subset \Xset $ are the \textit{centers}, and the
coefficient vectors $\kernelcoeff_i \in \setR^{\outDim}$ are determined by imposing interpolation conditions on $\Xset_{\bigN}$, i.e., \eqref{eq:kernel expansion} is
exact when evaluated at $\vrv \in X_N$. These conditions result in a linear system, which has a unique solution whenever the kernel function is chosen to be strictly
positive definite, i.e., for any choice of pairwise distinct points $\Xset_{\bigN}$ the matrix
$\kernelmatrix_{ij} :=
\kernel(\xcenters_i,\xcenters_j)$
is positive definite. Therefore, it is possible to construct an approximation \eqref{eq:kernel expansion} for arbitrary sample locations $\Xset_{\bigN}$ and input and
output
dimensions. In practice, orders of hundreds for input and output dimensions are realistic.

We use in the following a $C^2$ Wendland kernel \cite{Wendland1995a}, which is a compactly supported radial kernel of polynomial type, and where the radius of the
support is controlled by a \textit{shape parameter} $\kernelwidth>0$. Each kernel is associated to a \textit{native} Hilbert space $\RKHS(\Xset)$, which is in this case
a Sobolev space.

\begin{figure*}[htb]
\subfigure{\includegraphics[width=.48\textwidth]{\picdir/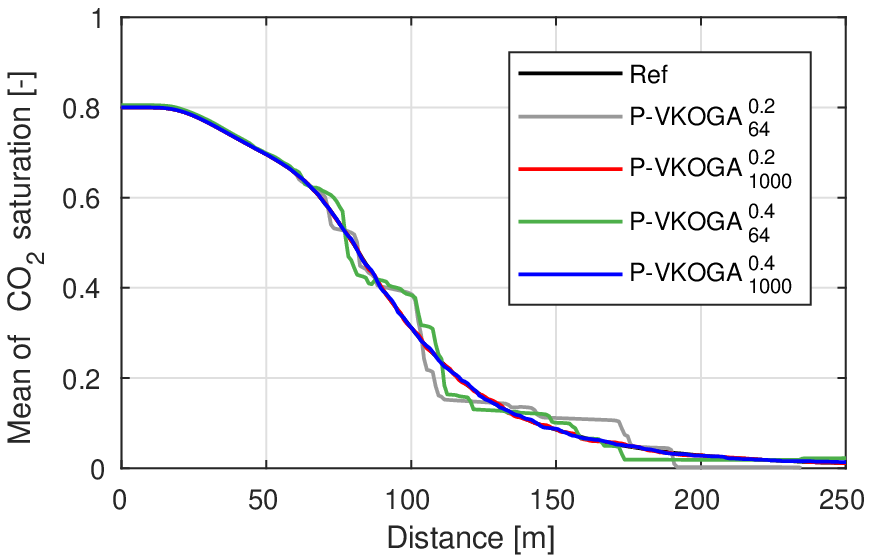}}\hfill
\subfigure{\includegraphics[width=.48\textwidth]{\picdir/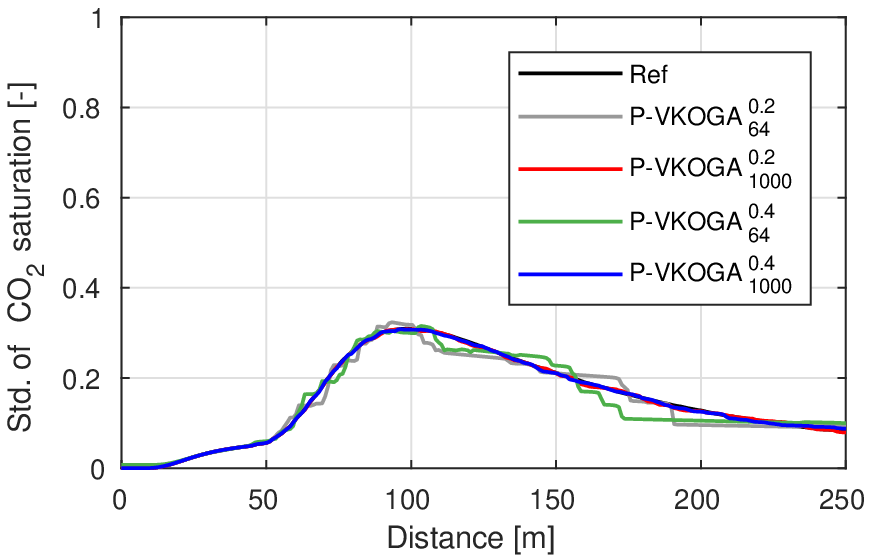}}
\caption{\label{fig:pgreedy_full} Mean and standard deviation of CO$_2$ saturation obtained via kernel interpolation with \vkoga{\smallN}{\kernelwidth} for the shape
parameters
$\kernelwidth=0.2$, $\kernelwidth=0.4$, and number of model runs $n = 64, 1000$.}
\end{figure*}

The quality of the kernel approximation in \eqref{eq:kernel expansion} depends on both the choice of $\kernel$ and the set of centers $\Xset_\bigN$. Given a large set of
possible
sample locations $\Xset_{\bigN}$, we want to select of a subset $\Xset_{\smallN} \subset \Xset_{\bigN}$ such that
the surrogate $\rf_n(\vrv)$ based on $X_n$ is as good as $\rf_N(\vrv)$, while ${\smallN} \ll {\bigN}$.
This ensures that the evaluation of the approximation \eqref{eq:kernel expansion} is as fast as possible, so that it can be efficiently used in surrogate modeling.

For this, we apply $\power$-VKOGA \cite{DeMarchi2005,Wirtz2013} (Vectorial Kernel Orthogonal Greedy Algorithm with $\power$-greedy),
to iteratively generate a nested sequence of centers $\Xset_{\smallN-1}\subset \Xset_{\smallN}\subset\Xset_{\bigN}$, $\Xset_0:=\emptyset$.
We give a rough outline for the motivation and structure of $\power$-VKOGA and refer to the aforementioned references for further details:
The interpolation error can be
bounded by means of the \textit{power function} $\power_{\Xset_{\smallN}} : \Xset \rightarrow [0,\infty)$ as
\begin{equation}
 \|\rf(\rd,t,\vrv) - \rf_n(\vrv)\|_{\infty} \leq \power_{\Xset_{\smallN}}(\kernelarg) \| \f \|_{\RKHS(\Xset)}\quad\forall \kernelarg\in \Xset
\label{eq: power function error estimate},
\end{equation}
where the norm on the left-hand side is the maximal absolute value of all entries of the
$d$-dimensional vector $\rf(\rd,t,\vrv) - \rf_n(\vrv)$.
Observe that, when the kernel $\kernel$ is understood as a covariance function, $\power_{\Xset_{\smallN}}$ is precisely the expected prediction error of the
corresponding \textit{simple Kriging}, as discussed in \cite{scheuerer2013}.

Motivated by \eqref{eq: power function error estimate} we chose the centers iteratively by adding $\kernelarg^{\ast} = \arg\max\limits_{\kernelarg \in
\Xset_{\bigN}\setminus
\Xset_{\smallN}}
P_{\Xset_{\smallN}}(\kernelarg)$
to the previous set of centers, i.e. $\Xset_{{\smallN}+1} := \Xset_{{\smallN}} \cup \{ \kernelarg^{\ast} \}$.
Using the Newton basis $\{v_i\}_{i=1}^{\smallN}$ \cite{Muller2009} of the space $\textrm{span}(\kernel(\cdot,\xcenters_i),\xcenters_i\in\Xset_{\smallN})$, the power
function can
be efficiently updated via $\power_{\Xset_{\smallN}}(\kernelarg)^2 = \kernel(\kernelarg,\kernelarg) - \sum\limits_{i=1}^{\smallN} v_i(\kernelarg)^2. \label{eq: power
function
representation}$ Once a sufficient set $\Xset_{\smallN}$ is generated, the coefficients $\kernelcoeff_i \in \setR^{\outDim}$ can be computed by solving the system
\eqref{eq:kernel
expansion} with interpolation conditions
restricted to the points $\Xset_{\smallN}$. This method requires only the knowledge of a sampling $X_N$ of the input space $\Xset$ to select the few sampling points
$\Xset_{\smallN}$, and the solver is run only on the parameters in $\Xset_{\smallN}$.
We remark that it has been recently proven that  $\power$-VKOGA, although simple, has a quasi-optimal convergence order in Sobolev spaces \cite{SH16b}, i.e., it gives
the same
approximation order of full interpolation.

In the present setting, the surrogate model is constructed on the set $\Xset$ defined as the convex hull of the reference data points $\dataPoints$. We run
$\power$-VKOGA starting from a
fine discretization $\Xset_{\bigN}$ of $\Xset$ obtained by intersecting $X$ with a uniform grid of $50\times50\times50$ equally spaced points in the minimum box
enclosing
$\dataPoints$. The resulting set $\Xset_{\bigN}$ contains $N = 86021$ data points. We remark again that this set is used to perform the greedy optimization, and no
function
evaluation (i.e., no model run) on $X_N$ is needed at this stage.
Since in this setting the number of model runs is the crucial computational constraint, we avoid to select the kernel width parameter $\kernelwidth$ via validation,
and instead
train different models for values $\kernelwidth = 0.1, 0.2, 0.3, 0.4, 0.5$.

Each run of the algorithm selects incremental sets $\Xset_{\smallN}$ of point locations, and for each of the $5$ parameters $\kernelwidth$ we consider $6$ log-spaced
values of
$\smallN$ in $[1, 1000]$, i.e., $\smallN = 1, 4, 16, 64, 252, 1000$. On these sets $\Xset_{\smallN}$ of input parameters, the full model is evaluated and for
each of them the interpolant is computed.
The resulting $30$ different models are denoted as \vkoga{\smallN}{\kernelwidth}, where the lower index is the number of model runs and the upper one is the kernel
shape
parameter. Sample results are shown in Fig.~\ref{fig:pgreedy_full} for values $\kernelwidth = 0.2, 0.4$ and for an increasing
number $n$. Although the surrogates are not accurate when using only $n=4$ model runs, it is evident that an increase in $n$ leads to a satisfactory
convergence of the approximate mean and standard deviation to the reference ones. Indeed, the surrogates obtained with $64$ model runs are close enough to the reference
solution, with some oscillations around the exact values.
For $1000$ runs, the reference and the surrogates are almost equal. A more quantitative error analysis is discussed in the next sections.

\subsection{Hybrid stochastic Galerkin}\label{sec:hsg}

Contrary to the previous methods, the hybrid stochastic Galerkin (HSG) approach is an intrusive method
which changes the deterministic system by means of the polynomial chaos expansion and a multi-element
decomposition of the stochastic space, hence, does not construct a surrogate.
We briefly summarize the discretization following
\cite{Buerger2012,Koeppel2017}. For the sake of brevity we again use $\Sat := \eSa\nw$.

If all sources of uncertainty are considered, c.f. Section \ref{sec:model}, the radial transport equation
\eqref{2phase_fracflow_rad_2} yields the following randomized partial differential equation
\begin{equation}\label{eq:2phase_random}\displaystyle
\pore(\vrv)\frac{\partial\Sat(\vrv)}{\partial t} - \frac{1}{\rd} \frac{\partial}{\partial \rd}
\big(\urate(\vrv)\fxg(\Sat,\vrv) \big) - \inRate = 0.
\end{equation}

For the multi-element discretization we decompose the stochastic domain 
into $2^{\sdim\Nr}$ stochastic elements with refinement level $\Nr\in\setN_0$.
For reasons of readability we describe the discretization based on the domain $[0,1]^\sdim$. 
By rescaling, the method can be easily extended to arbitrary domains.
Let $\setIt:=\{0,1,\ldots,2^{\Nr}-1\}$ be a set of indices and
$\setI:=\setIt^{\sdim}$ with multi-index $\mi{l}=(\smi{l}{1},\ldots,\smi{l}{\sdim})\in\setI$.
Then we define the $\sdim$-dimensional stochastic element
$\elem{\sdim}{\mi{l}}:=\elemA{\smi{l}{1}}\,\times\ldots\times\,\elemA{\smi{l}{\sdim}}$ with support
$\elemA{\smi{l}{i}}:=[2^{-\Nr}\smi{l}{i},\,2^{-\Nr}(\smi{l}{i}+1)]$, for $\smi{l}{i}\in\setIt$,
$i\,=\,1,\dots,\sdim$.
Furthermore let $\SE{\sdim}{\No}{\Nr}$ be the space of piecewise polynomial functions on each
stochastic element $\elem{\sdim}{\mi{l}}$ with maximal polynomial order $\No$.
The space $\SE{\sdim}{\No}{\Nr}$ is spanned by the multivariate polynomials
\begin{equation*}\label{eq:mvHSGpoly}
 \BpolMR[\Nr]{\mi{p}}{\mi{l}}(\vrv):=\begin{cases}
  2^{\sdim\Nr /2}\prod_{k=1}^{\sdim} \bpol{\mi{p}_k}(2^{\Nr}\rvi{k}-\smi{l}{k}), &
  \mbox{for}\;\vrv\in\elem{\sdim}{\mi{l}},\\
  0,&\text{otherwise},
  \end{cases}
\end{equation*}
with $\mi{p}:=(\mi{p}_1,\ldots,\mi{p}_\sdim)\in\setN_0^{\sdim},\;\abs{\mi{p}}\leq\No$, $\mi{l}\in\setI$, and
the truncated polynomial chaos polynomials $\bpol{\mi{p}_k}(\rvi{k})$. For the latter we use Legendre polynomials.
The polynomials $\BpolMR[\Nr]{\mi{p}}{\mi{l}}(\vrv)$ satisfy the following orthonormality relation
$
 \prdV{\BpolMR[\Nr]{\mi{p}}{\mi{l}}}{\BpolMR[\Nr]{\mi{q}}{\mi{m}}}:=
 \delta_{\mi{p},\mi{q}}\delta_{\mi{l},\mi{m}},
$
where $\delta_{\mi{p},\mi{q}},\,\delta_{\mi{l},\mi{m}}$ denote the Kronecker delta symbol for
$\mi{p},\mi{q}\in\setN_0^\sdim,\;\abs{\mi{p}},\abs{\mi{q}}\leq\No$, and $\mi{l},\mi{m}\in\setI$.
The finite number of basis functions is given by $\nP :=2^{\sdim\Nr}(\No+\sdim)!/(\No!\sdim!)$.
Based on these considerations the projection of a random field $\rf(\rd,t,\vrv)$, $(\rd,t)\in\setD_T$
is obtained by
\begin{multline}\label{eq:mvHSGprj}
 \rf(\rd,t,\vrv)\approx\,\Prj[\rf]{\No}{\Nr}(\rd,t,\vrv) \\ :=  \sum_{\mi{l}\in\setI} \sum_{p=0}^{\No} \sum_{\abs{\mi{p}}=p}
 \mrcc[\rf]{\Nr}{\mi{p}}{\mi{l}}(\rd,t)\BpolMR[\Nr]{\mi{p}}{\mi{l}}(\vrv), \qquad
\end{multline}
with deterministic coefficients $\mrcc[\rf]{\Nr}{\mi{p}}{\mi{l}}:=\prdV{\rf}{\BpolMR[\Nr]{\mi{p}}{\mi{l}}}$,
for $0\leq\abs{\mi{p}}\leq\No$ and $\mi{l}\in\setI$.
We note that without the multi-element decomposition the expansion \eqref{eq:mvHSGprj} would be similar to \eqref{PCE}.
For more details concerning the convergence of
$\Prj[\rf]{\No}{\Nr}$ for $\No,\,\Nr\to\infty$, we refer to \cite{MR1199538}.
Moreover, we refer to \cite{Buerger2012,Koeppel2017} for a more detailed description of the HSG method.

We apply the stochastic discretization to \eqref{eq:2phase_random}
by replacing the unknown random field, i.e. the saturation $\Sat(\x,t,\vrv)$, with its
HSG projection $\Prj[\Sat]{\No}{\Nr}$ and by testing the equation with the HSG basis functions
$\BpolMR[\Nr]{\mi{p}}{\mi{l}}$. Then we obtain a partly decoupled system for the deterministic coefficients
$\mrcc[\Sat]{\Nr}{\mi{p}}{\mi{l}}$ which reads
\begin{multline} \label{eq:ds2pm.h}
 \partial_t \mrcc[\Sat]{\Nr}{\mi{p}}{\mi{l}} + (1/r)\;\divg \prdV{\urate\fxg\big(\Prj[\Sat]{\No}{\Nr}\big)/\pore}
 {\BpolMR[\Nr]{\mi{p}}{\mi{l}}} \\
 - \prdV{\inRate/\pore}{\BpolMR[\Nr]{\mi{p}}{\mi{l}}} = 0, \quad
\end{multline}
with initial values $\mrcc[\Sat]{\Nr}{\mi{p}}{\mi{l}}(\cdot,0)=\prdV{\Sat_0}{\BpolMR[\Nr]{\mi{p}}{\mi{l}}}$.
On $\partial\setD$ we impose deterministic boundary conditions.

\begin{figure*}[htb]
\subfigure{\includegraphics[width=.48\textwidth]{\picdir/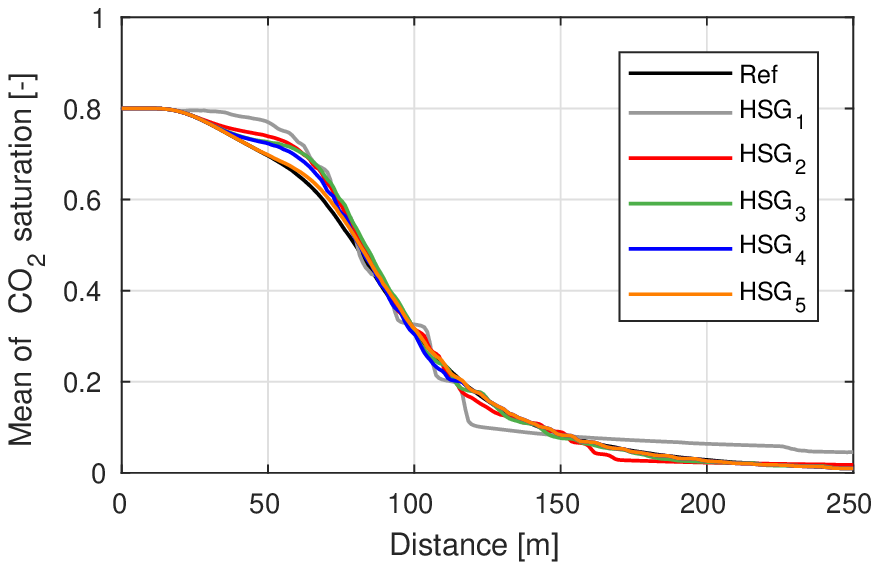}}\hfill
\subfigure{\includegraphics[width=.48\textwidth]{\picdir/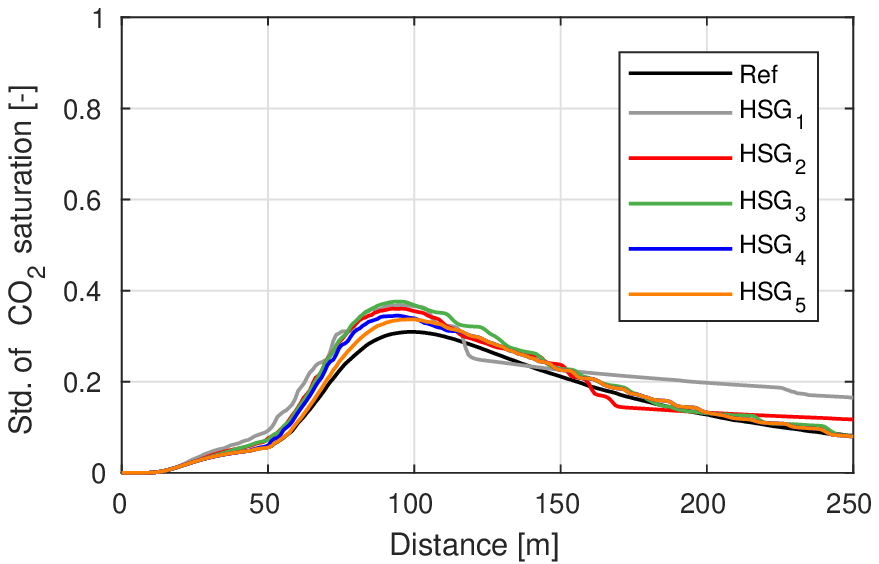}}
\caption{\label{fig:HSG_full} Mean and standard deviation of CO$_2$ saturation obtained via reconstruction of HSG discretization wihtout and with decomposed stochastic space.
HSG$_1$: $\Nr=2$, $\No=1$.
HSG$_2$: $\Nr=3$, $\No=1$.
HSG$_3$: $\Nr=2$, $\No=2$, and $\Nr=3$, $\No=3$.
HSG$_4$: $\Nr=3$, $\No=1$, and $\Nr=3$, $\No=1$.
HSG$_5$: $\Nr=3$, $\No=1$, and $\Nr=4$, $\No=1$.
}
\end{figure*}

Mean and standard deviation are then computed during the post-processing.
In contrast to the common SG/HSG ansatz, in this work the simulations are performed with
Legendre polynomials representing a uniform distribution even if the actual distribution
might be different. In view of data-driven UQ this guarantees that every point is uniformly approximated.
Instead of using the coefficients of \eqref{eq:mvHSGprj} directly mean and standard deviation
are obtained by a reconstruction using the parameters provided in the data set $\dataPoints$.
This is possible since equation~\eqref{eq:mvHSGprj} allows to compute the approximation of
the unknown random field for an arbitrary parameter set. However this approach is likely to be
at the expense of the accuracy. On the other hand we obtain mean and standard deviation for
arbitrary input data without a rerun of the simulation.

Fig.~\ref{fig:HSG_full} shows the results of the HSG method compared with the reference solution.
An increase of $\Nr$ leads to a greater improvement of the results than an increase of $\No$ only.
Since in the test case the value of the perturbed porosity $\pore$ has a significant
influence on the dynamics of the system, some of the shown simulations are decoupled
in two parts w.r.t. the value of $\pore$. By merging the decoupled results during the
post-processing the accuracy can be improved.
Even though the time steps on different stochastic elements differ from each other, the time step size on each
stochastic element of the considered problem does not change during the whole computation time. Therefore, we 
improved the computational efficiency significantly by rearranging the stochastic elements on several Message
Passing Interface (MPI) ranks in terms of the expected number of time steps.



\section{Discussion}\label{sec:discussion}
This section discusses the results obtained by the four different methods
in more detail and compares them with respect to five different criteria. The results are visualized in Fig.~\ref{fig:L2_norm_joint_loglog} to facilitate the reading.
Moreover we report in Table~\ref{tab:compariso_table} a
summary of the discussion to provide the reader with an immediate comparison of the strengths and weaknesses of the different techniques.
However we emphasize that the comparison needs to be evaluated with caution due to the inherent differences of the methods.

\subsection{Usage for UQ in geosciences}\label{sec:discussion-used-in-UQ}
The polynomial chaos expansion is a known technique in the field of uncertainty quantification and its generalization to arbitrary distributions has been applied already to CO$_2$ storage problems. Similarly, spatially adaptive sparse grids are a common stochastic collocation method in uncertainty quantification that has been applied to a large variety of real-world problems. Kernel-based methods are successfully applied in a variety of data-based tasks, in particular in function approximation, but still novel in uncertainty quantification in geosciences. The use of kernel models in the context of uncertainty quantification in principle poses no particular issues, as the mean and variance predictions can be simply obtained by evaluating the surrogate model and computing the corresponding empirical mean and variance.

By construction, intrusive stochastic Galerkin methods change the structure of the problem. Consequently,
the quantification of uncertainties goes along with the underlying deterministic problem to be solved.
In the early 90's, intrusive stochastic Galerkin methods have been successfully used for the quantification of uncertainties of elliptic problems, e.g. by Ghanem and Spanos
\cite{Ghanem_Spanos_1991_SFEM_book}.
In the last decades these methods have been extended further to cope with several applications of different complexity. In particular, nonlinear hyperbolic problems require an
additional stochastic discretization, such as the multi-wavelet approach or the HSG method, involving the decomposition of the stochastic space in order to avoid or reduce
the generally occurring Gibbs phenomenon.

\subsection{Comput. costs to evaluate \& reconstruct surrogate}\label{sec:discussion-comput-costs}
Polynomial chaos expansions and its recent data-driven aPC generalization can be seen as a cheap way for the estimation of uncertainty. Therefore the computational
costs involved are comparatively low.

Due to its nature, the same applies to sparse grid methods.
All sparse grid methods used in this paper are publicly
available in SG$^{\mathrm{++}}$ \cite{Pflueger10Spatially}. This
toolbox includes state-of-the-art sparse-grid algorithms that allow for
very efficient and parallel construction, evaluation and quadrature of
sparse grid surrogates.

Greedy methods prove to be particularly effective in general surrogate modeling since they produce sparse, hence cheap-to-evaluate models. This is the case also in the present setting,
while greedy data-independent methods, as the presently used $P$-greedy algorithm, have the further advantage of not requiring an a-priori knowledge of the full model evaluations.
Indeed, one of the measures for the success of a method is the number of model runs required to construct the surrogate.

The results of the HSG method with low resolution can be computed with appropriate computational effort, but the complexity of the problem and also the computational costs
increase
rapidly with increasing resolution. However, the method facilitates a change of the probability distribution provided that it is still defined on the same
interval, almost without any computational costs during the postprocessing. Furthermore the partially decoupled structure of the HSG-discretized  problem allows for efficient
parallelization on MPI and Open Multi-Processing clusters. Additional improvements of the computational efficiency can be achieved by the stochastic adaptivity~\cite{Barth201611,Buerger2012}
and load balancing based on the stochastic elements.
It is also promising to exploit the vector structure of the discretization by using GPU-based architectures.

\subsection{Accuracy for low number of model runs/resolution}\label{sec:discussion-accuracy-low}
Low-order aPC representations such as 1$^{st}$ and 2$^{nd}$ order seem to be efficient in terms of computational costs and corresponding accuracy (see
Fig.~\ref{fig:L2_norm_joint_loglog}). However, the aPC approach covers the parameter space globally and as a consequence has not enough flexibility to represent special
features in parameter space such as strong discontinuities, shocks, etc.

As shown in Fig.~\ref{fig:L2_norm_joint_loglog},
sparse grids are efficient with respect to accuracy and the
corresponding computational costs. For a very limited computational
budget with less than $20$ model runs, the convergence of the sparse
grid with the modified basis is limited by the grid structure from
which the interpolation points are taken.

The results of the $P$-greedy method show that  a certain minimal number of data is required to have meaningful predictions (at least $50$ model runs). This is motivated by two
factors: first, the model does not incorporate any knowledge on the input space distribution. Second, the $P$-greedy algorithm selects the input points in a
target-function-independent way, so it is not specialized on the approximation of this specific model.

The HSG method provides suitable accuracy for low resolutions of mean and standard deviation, cf. Fig.~\ref{fig:L2_norm_joint_loglog}. This applies in particular to the bisection
level $\Nr$, i.e. the number of stochastic elements. The choice of the polynomial order $\No$ has no significant bearing on the accuracy of the considered problem.

\subsection{Accuracy for high numb. of model runs/resolution}\label{sec:discussion-accuracy-high}
\begin{figure*}
\subfigure{\includegraphics[width=.5\textwidth]{\picdir/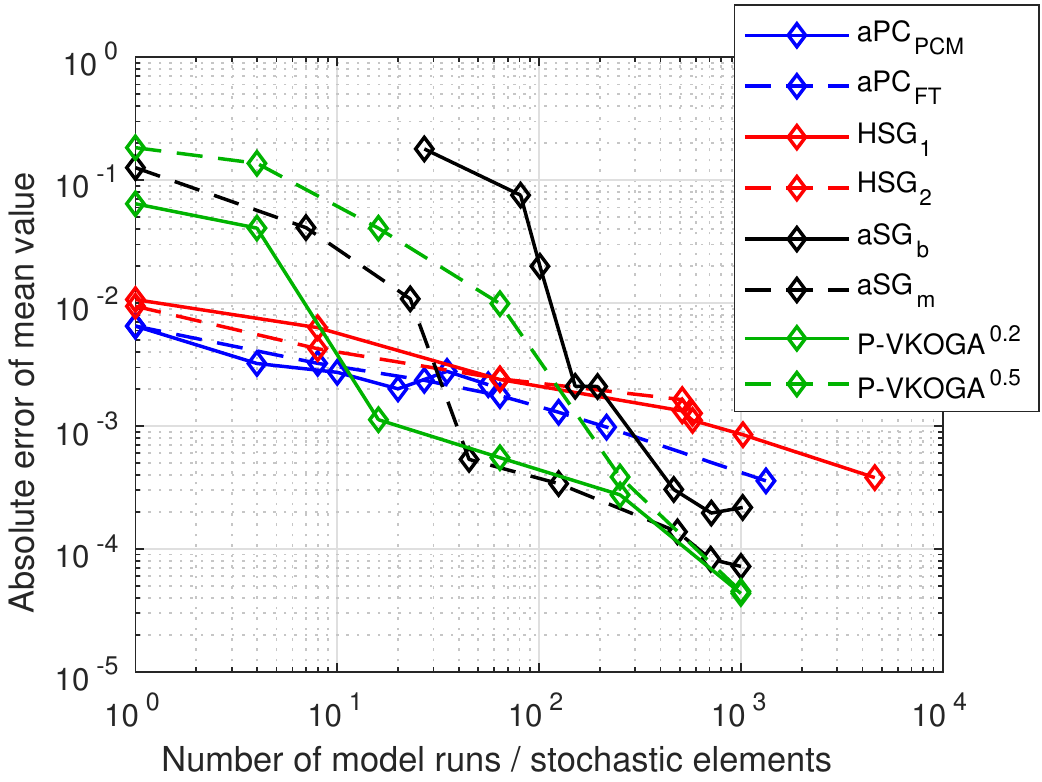}}
\subfigure{\includegraphics[width=.5\textwidth]{\picdir/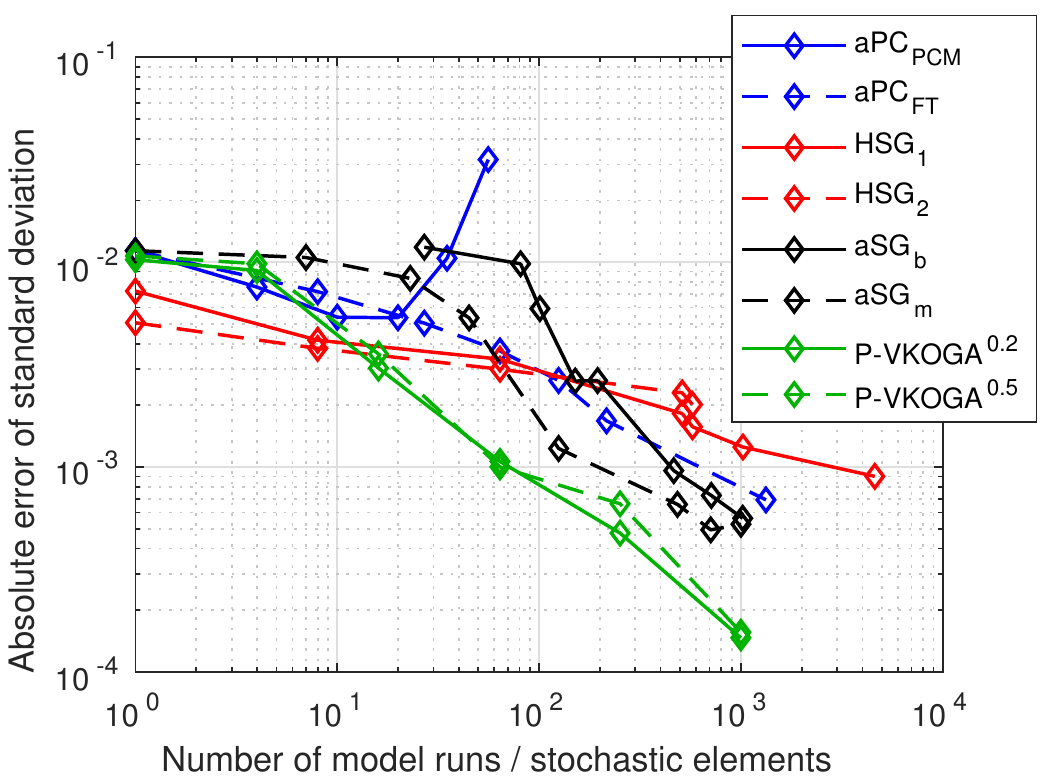}}
\caption{\label{fig:L2_norm_joint_loglog} Absolute error of mean and standard deviation versus the number of model runs / stochastic elements (cf. HSG): 
aPC$_{PCM}$ - aPC expansion based on the probabilistic collocation method; aPC$_{FT}$ - aPC expansion based on the least-squares collocation method;
HSG$_1$ - HSG method with maximal polynomial order $\No=1$;
HSG$_{2}$ - HSG method with maximal polynomial order $\No=2$;
\asgbs - adaptive sparse grids with boundary points;
\asgms - adaptive sparse grids without boundary points but with a modified basis;
 $\power$-VKOGA$^\kernelwidth$ - kernel greedy interpolation with shape
parameter $\kernelwidth = 0.2, 0.5$.
}
\end{figure*}

As pointed out in Section~\ref{sec:nipc}, an increase of the expansion order in aPC does not necessarily lead to an improvement. This fact is very well illustrated in
Fig.~\ref{fig:L2_norm_joint_loglog}. Apparently the 0$^{th}$ order estimation of the standard deviation which suggests by definition a value of $0$ and demands one run of the original model only, is more accurate than the 5$^{th}$ order expansion. Alternatively, this artifact can be mitigated via least-squares projection onto the full tensor grid, as presented in Fig.~\ref{fig:L2_norm_joint_loglog}. The aPC based on least-squares collocation helps to overcome the problem of representing discontinuities and assures the reduction of the  error. However this is at the expense of the computational costs involved.

The efficiency of the sparse grids method with modified basis increases
significantly when the adaptive refinement comes into play, as both the level of the
grid and the accuracy of the surrogate increase in the regions
of high probability. The error keeps converging as expected with an
increasing number of model runs.
Sparse grids with boundary points converge as well, however with
higher computational costs (see Fig.~\ref{fig:L2_norm_joint_loglog}).
Consequently, they should not be considered for data-driven uncertainty
quantification problems in higher dimensions or with small computational
budgets.

The results and comparisons of Fig.~\ref{fig:L2_norm_joint_loglog} demonstrate a good behavior of the $P$-greedy algorithm in the present task and, in particular, suitable numerical
convergence as
the size of the dataset increases. This behavior is particularly evident for larger datasets
(i.e. when at least $50$  model runs are utilized), as the surrogate model improves its accuracy, which turns out to be an advantage over other methods.

The HSG method provides fair accuracy for high resolutions of mean and standard deviation
for increasing $\Nr$ and $\No$. An increase of the bisection level $\Nr$ usually leads to higher accuracy than
the increase of the highest polynomial order $\No$, cf. Fig.~\ref{fig:L2_norm_joint_loglog}.
Because of the considered setup, numerical quadrature needs to be performed in each time step.
Depending on the desired accuracy, the discretization of the setup includes up to several hundreds of thousands time steps. This may result in numerical difficulties and also
accuracy limitations.

\subsection{Applicability for large number of parameters}\label{sec:discussion-high-dimensions}
In general, the different aPC approaches used in this work are not very suitable for high-dimensional problems due to the curse of dimensionality. Only an extension to
sparse polynomial representation can be feasible for applied tasks \cite{ahlfeld2016samba}.

In contrast, spatially adaptive sparse grids have three main advantages compared to the other approaches.
First, as pointed out in Section~\ref{sec:sgrid},
they are suited for higher-dimensional problems. Second, they
adaptively allow to change the approximation locally wherever local features in
the parameter space appear. Large hierarchical coefficients describe
large local changes and, hence, serve as a basis for refinement
criteria. It is even possible to cope with functions that include
kinks or jumps. And third, they are very flexible and allow
$p$-adaptive refinement. This means one can choose the polynomial degree of
each basis function separately in order to exploit local smoothness of
the model function.

The $P$-greedy method can potentially work on problems with many more input dimensions, although a slower convergence rate should be
expected due to the curse of dimensionality. Nevertheless, the complexity of the model construction and evaluation remain essentially the same, up to the computation of Euclidean
distances between points in a larger input space.

As mentioned, all intrusive techniques base on a transformation of the randomized problem into a deterministic system of equations and, at least in the general case,
they
require the application of a quadrature in each time step.
Compared to non-intrusive methods this methodology therefore changes the structure of the problem.
Consequently, intrusive methods including HSG suffer from the curse of the dimensionality.

\begin{table*}[ht]
\begin{center}
\begin{tabular}{|l|>{\centering}m{2.35cm}|>{\centering}m{2.55cm}|>{\centering}m{2.8cm}|>{\centering}m{2.8cm}|>{\centering\arraybackslash}m{2.7cm}|}
\hline
&Usage in geoscience UQ&Comput. costs to reconstruct \& evaluate surrogate &Accuracy for low number of model runs/resolution
&Accuracy for high number of model runs/resolution&High-dimensional applicability\\
\hline
aPC$_{PCM}$&\goodFirstCol&\good&\good&\bad&\bad\\
\hline
aPC$_{FT}$&\okFirstCol&\good&\good&\ok&\bad\\
\hline
aSG&\okFirstCol&\good&\ok&\ok&\ok\\
\hline
$\power$-VKOGA&\badFirstCol&\good&\ok&\good&\ok\\
\hline
HSG&\goodFirstCol&\ok&\good&\ok&\bad\\
\hline
\end{tabular}
\caption{Comparison of the methods w.r.t. the five criteria of Section~\ref{sec:discussion}.
The different colors reflect the behavior of the methods 
}
\label{tab:compariso_table}
\end{center}
\end{table*}


\section{Summary and Conclusions}\label{sec:sum}
In this benchmark study we have compared four promising techniques 
for data-driven uncertainty quantification of nonlinear two-phase flow in porous media.
The flow problem is motivated by an injection scenario of supercritical $\coo$ into
a porous and saline aquifer where the consideration of uncertainty in the boundary
conditions, uncertainty in material parameters as well as conceptual model uncertainty
has a significant bearing on the overall flow and storage behavior of the system.
Therefore, we have considered uncertainty in the injection rate related to boundary value uncertainty, uncertain porosity related to the limited
data availablity of field-scale storage sites and uncertainty of the relative permeability degree associated with the nonlinearity of the conceptual model.
To account for the arising uncertainties, we considered
the non-intrusive methods arbitrary polynomial chaos, spatially adaptive sparse grids
and kernel greedy interpolation as well as the intrusive hybrid stochastic Galerkin
method. They were compared by means of the absolute error of the
moments mean and standard deviation of the $\coo$ saturation after 100 days based on
a statistical reference solution.

The numerical results show that all methods provide a good representation of the
considered moments of the quantity of interest, despite the inherent complexity
stemming from the distribution and the impact of the uncertain parameters. Small changes of these
parameters may strongly influence the nonlinearity of the flow problem
and may cause large variability of the simulation time of the hyperbolic solver
resulting in significant changes of the $\coo$ saturation and the shape of the plume.
This usually leads to Gibbs phenomena and oscillations of response surfaces in UQ methods.
The arbitary polynomial chaos method and the hybrid stochastic Galerkin method
overcome these difficulties already for low number of model runs/resolution, 
whereas adaptive sparse grids and the kernel greedy method are characterized by improved accuracy at
higher number of model runs/resolution.
The applicability to high dimensionalities
needs to be taken into account. Our discussion indicates that the arbitary polynomial
chaos method and the hybrid stochastic Galerkin method are generally less efficient
and typically suffer more from the curse of dimensionality compared to adaptive sparse
grids and the kernel greedy interpolation technique.
The particular method of choice therefore depends on the specific problem and specific goal to be achieved.
Taking this fact into consideration we have classified the methods regarding five relevant properties which are presented in Table~\ref{tab:compariso_table}. The classification reflects the main features 
we deem to be useful for further uncertainty quantification for $\coo$ storage and beyond.

\begin{acknowledgement}
The authors would like to thank the German Research Foundation (DFG) for financial support of the project within the Cluster of Excellence in Simulation Technology (EXC 310/2) at
the University of Stuttgart. 
\end{acknowledgement}

\bibliographystyle{spmpsci}      
\bibliography{lit}
\end{document}